# Tunable Fluorescein-Encapsulated Zeolitic Imidazolate Framework-8 Nanoparticles for Solid-State Lighting

Tao Xiong,[a,†] Yang Zhang,[a,†] Lorenzo Donà,[b] Mario Gutiérrez,[a] Annika F. Möslein,[a]

Arun S. Babal,[a] Nader Amin,[c] Bartolomeo Civalleri,[b] and Jin-Chong Tan[*,a]

[a]Multifunctional Materials & Composites (MMC) Laboratory, Department of Engineering Science, University of Oxford, Parks Road, Oxford OX1 3PJ, UK.

[b]Department of Chemistry, NIS and INSTM Reference Centre, University of Turin, via Pietro Giuria 7, Torino 10125, Italy.

[c]Department of Chemistry, University of Oxford, Mansfield Road, Oxford OX1 3TA, UK

[†]These authors contributed equally to this work

*Corresponding author: jin-chong.tan@eng.ox.ac.uk

**Abstract**

A series of fluorescein-encapsulated zeolitic imidazolate framework-8 (fluorescein@ZIF-8) luminescent nanoparticles with a scalable guest loading has been fabricated and characterized. The successful encapsulation of the organic dye (fluorescein) is supported by both experimental evidence and theoretical simulations. The measured optical band gap is found to be comparable with the computed values of a hypothetical guest-host system. Isolated monomers and aggregates species of fluorescein confined in ZIF-8 nanocrystals have been systematically investigated through fluorescence lifetime spectroscopy. The quantum yield (QY) of the obtained solid-state materials is particularly high (QY~98%), especially when the concentration of the fluorescein guest is low. Combining a blue LED chip and a thin photoactive film of fluorescein@ZIF-8, we demonstrate a device with good optical tunability for multicolor and white light emissions. Additionally, we show that the fluorescein@ZIF-8 nanoparticles exhibit an improved photostability due to the shielding effect conferred by the nanoconfinement of host framework, making them promising candidates for practical applications such as solid-state lighting, photonics, and optical communications.

**Introduction**

Luminescent materials are widely employed in a vast range of technologies encompassing white-light-emitting diodes (WLEDs), optoelectronics, fluorescent microscopy and bioimaging, optical sensors, and telecommunications. In terms of solid-state lighting, several approaches to implement WLEDs are being investigated. In one approach, three independent light-emitting diodes (LEDs), one for red, green, and blue each, are combined to produce a white light.[1] Although conceptually simple, this color-mixing method is practically complicated in that additional circuitry is needed to coordinate the three-component LEDs as their working voltages and lifespans, for example, may be different, thereby imposing an extra burden on cost. By far a more popular and successful approach involves a combination of a blue LED chip and a phosphor which can be excited to emit either a wide spectrum ranging from the red to the green or a relatively narrow spectrum focused in the yellow. Part of the blue light from the source LED is absorbed by the phosphor, whose emission spectrum is mixed with the rest of the blue light to render a white light. Apart from being much more energy efficient than conventional technologies, which is common to all WLEDs, this so-called (partial) phosphor-conversion approach[2] has the added advantage of being more tunable, usually by manipulating the conditions under which the phosphors are prepared.

Early research on energy efficient phosphors typically involves rare-earth elements (REEs).[3,4] This raises economic, environmental, and geopolitical concerns due to the increasing restrictions on the supply end of REEs and consequent surging prices, the hostility of the production of REEs to the environment, and the uneven geographical distribution of REEs and the associated governmental strategic monopoly.[5] These considerations together with the growing market demand for REEs in the foreseeable future have stimulated an ongoing interest and presented challenges for researchers to develop REE-free phosphors with a better energy efficiency. Notable progress has been achieved towards this goal with the development of various categories of inorganic phosphors.[6-10] Organic dye containing materials as potential phosphors for WLED applications are also being explored.[11-15] Due to the relatively limited success of efficient WLEDs based on organic phosphors in the market and increasing market demand for cost-effective WLEDs, further research in this field is needed.

Generally, the organic fluorophores (dye molecules) exhibit enhanced luminescent properties (in terms of a higher quantum yield) when isolated in a highly diluted solution state. However, most of them suffer from aggregation-induced quenching in the solid state, hindering their potential application in WLEDs and solid-state photonics. Furthermore, the fluorophores might not possess long-term photostability to deliver reliable performance under the harsh conditions of an operating blue LED. One efficient solution to surpass these drawbacks is to disperse the dyes in a protective host environment. It was not until the recent decade when metal-organic frameworks (MOFs) were introduced to the WLED field, drawing the attention of interdisciplinary research to re-evaluate the promising potential of employing organic dyes as an efficient phosphor for WLEDs.[14-17]

MOFs are a class of nanoporous framework materials with versatile potential applications in sensors,[18-20] dielectrics and optoelectronics,[21,22] catalysis,[23,24] gas storage,[25] and biomedicine.[26,27] Their open framework structure can act as a 'host' to incarcerate many types of 'guest' including organic dye molecules by means of 'guest@host' nanoscale confinement.[28,29] By careful and rational selection of the dye and MOF, dye@MOF systems with a comparable or even higher quantum efficiency than conventional phosphor materials can be synthesized. Furthermore, the host framework structure could provide extra protection from photodegradation, thus enhancing the overall stability of the encapsulated dyes.

Fluorescein is a well-known organic dye molecule whose fluorescent quantum yield (QY) in diluted aqueous solutions can be very high (fluorescein dianion QY ~93%).[30,31] It is widely used as a fluorescent coloring agent for medical diagnostic imaging and for membrane and protein labelling.[32] The emission of fluorescein falls in the green-yellow region, making it a natural candidate for WLED yellow phosphors. Some of us have shown that fluorescein (a green-/yellow-ish emitter) and rhodamine B (a red emitter) together constitute a yellow phosphor in the context of a dual-guest@ZIF-8 system,[13] yielding a QY value of ~47% in the solid state. Although a single-guest fluorescein@ZIF-8 system has been reported by Zhuang *et al*.[26] in the context of a small drug molecule capture and delivery system, it deserves a much deeper investigation for better understanding of the underlying photophysics and guest-host interactions, improving the QY, and assessing its applicability to WLEDs and solid-state photonics.

In this work, we investigate the detailed photophysical and photochemical properties of a series of dye@MOF material comprising fluorescein encapsulated in zeolitic imidazolate framework-8 (fluorescein@ZIF-8), where the host is a MOF with tetrahedrally coordinated zinc ($ZnN_4$) bridged by 2-methylimidazole (mIm) ligands to form sodalite cages. The size (largest interatomic distance, including *van der Waals* radii) of fluorescein is 1.27 nm (according to our computation, see Computational Details), while the minimum/maximum dimension of the internal cage of ZIF-8 is 1.16/1.34 nm, which is sufficiently large to accommodate a fluorescein molecule (if it adopts a suitable orientation) while preventing aggregation in the same cage, thereby helping to spatially disperse the guest molecules and mimicking the environment of fluorescein in dilute solutions. We begin by synthesizing a variety of fluorescein@ZIF-8 samples using different concentrations of fluorescein solution in methanol (MeOH). The 'high-concentration reaction' (HCR) method[14,33,34] was employed to achieve a rapid one-pot and *in situ* encapsulation of fluorescein in ZIF-8. The fluorescein@ZIF-8 materials are characterized using atomic force microscopy (AFM), powder X-ray diffraction (PXRD), and attenuated total reflectance Fourier-transform infrared spectroscopy (ATR-FTIR). The infrared (IR) spectra of the guest, host, and guest-host systems have also been calculated using *ab initio* density functional theory (DFT) to help with the interpretation of the experimental spectra and to seek evidence of the nanoconfinement of the fluorescein guest within the pores of ZIF-8. Photoluminescent properties are systematically measured through excitation-emission studies combined with quantum yield and decay lifetime experiments. Finally, the potential applications of the fluorescein@ZIF-8 materials to WLEDs are demonstrated.

**Results and discussion**

**Synthesis and structure of fluorescein@ZIF-8**

The fluorescein@ZIF-8 samples (henceforth abbreviated as fluo@ZIF-8, Figure 1a) were synthesized by adopting the HCR method.[14,33,34] Briefly, two methanolic solutions containing fluorescein and $Zn(NO_3)_2 \cdot 6H_2O$, respectively, were combined with another methanolic solution of 2-methylimidazole and triethylamine ($Et_3N$) at room temperature (see the *Experimental Section*). A yellowish to orange solid product (whose color varies depending on the concentration of the fluorescein used) formed instantaneously. The crystal size was characterized by AFM (Figure 1b) and found to be nanoparticles in the range of *ca*. 50-100 nm, resembling the morphology of a previously reported dye@MOF material termed Gaq3@ZIF-8.[14] For comparison, the pristine ZIF-8 was also synthesized using two methods: the rapid HCR method and the conventional 'slow' method (24 h, without $Et_3N$).

The crystalline structure of the solid samples was characterized by PXRD (Figure 1c). The PXRD patterns of fluo@ZIF-8 of different concentrations match the simulated and measured patterns of pristine ZIF-8, providing evidence that the ZIF-8 host structure has formed successfully and retained upon fluorescein encapsulation in all the samples. The broadening of the Bragg peaks evidenced in Figure 1c, for samples synthesized using the rapid HCR method compared with ZIF-8 obtained by slow synthesis, can be attributed to the fine crystalline size.[35] Solution $^1$H NMR spectroscopy was employed to characterize the amount of guest loading as shown in Figure 1d, revealing a strong correlation to the amount of fluorescein dye introduced during the synthesis step. However, the location of fluorescein guests (either encapsulated or attached on the surface) cannot be determined directly from PXRD or NMR, therefore further investigations using FTIR and fluorescence spectroscopy will be needed.

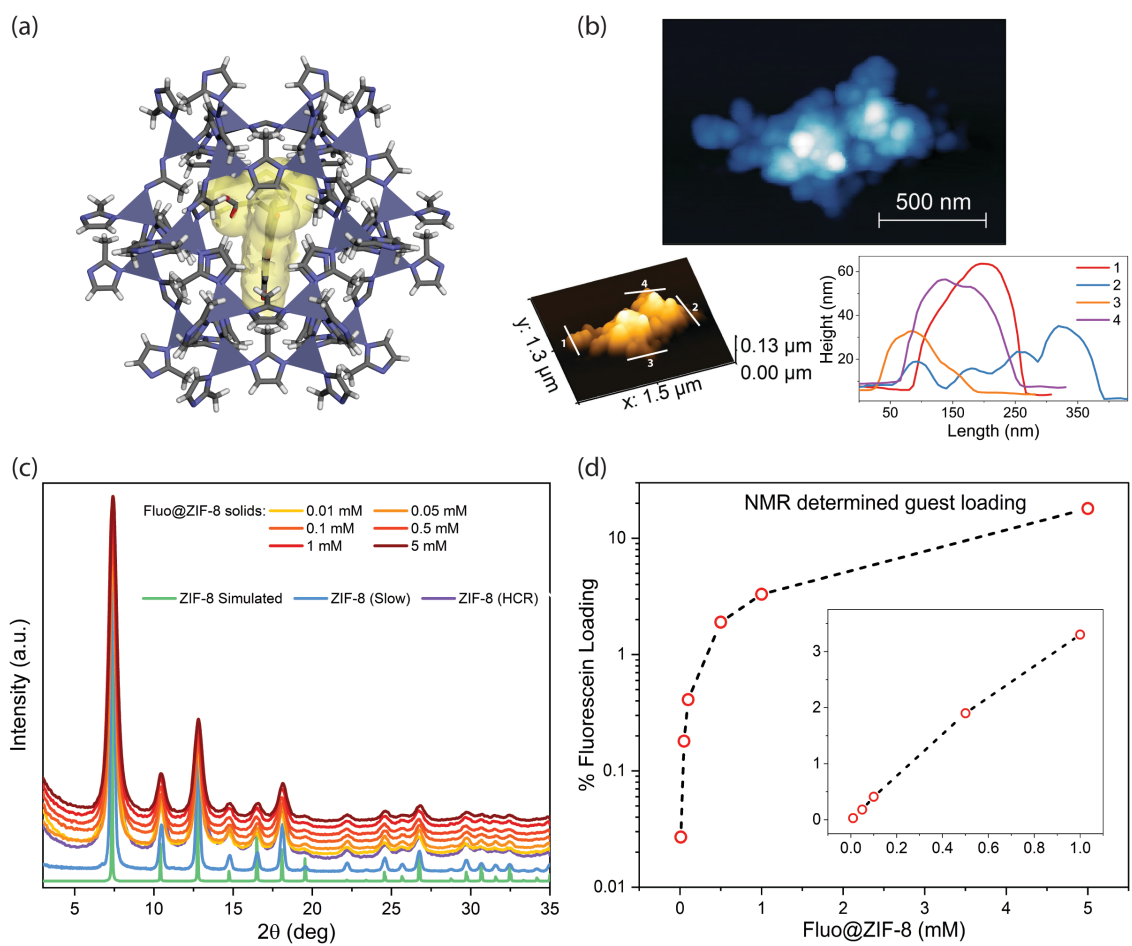

**Figure 1.** (a) Schematic representation of a fluorescein guest molecule encapsulated inside the pore of a ZIF-8 sodalite cage, to yield the fluo@ZIF-8 composite system. The purple tetrahedra are ZnN$_4$. Color code of atoms - red: oxygen, gray: carbon, blue: nitrogen, white: hydrogen. (b) AFM image of the fluo@ZIF-8 nanocrystals and the corresponding height topography extracted from the designated paths on the 3D surface reconstruction. (c) PXRD patterns of the simulated and synthesized ZIF-8 (HCR and 'slow' methods) and the six synthesized samples of fluo@ZIF-8 featuring six different fluorescein concentrations (mM) applied during material synthesis. The simulated pattern of ZIF-8 is calculated using the crystallographic information file (CIF) obtained from the Cambridge Structural Database (CCDC code: VELVOY). (d) Fluorescein guest loadings determined by solution $^1$H NMR spectroscopy. Inset shows the loading amount plotted in linear scale. The dashed lines are guides for the eye.

**Infrared spectra of fluorescein@ZIF-8**

To shed light on the composition of fluo@ZIF-8, the ATR-FTIR spectra of the synthesized fluo@ZIF-8 samples are measured and compared to those of pristine ZIF-8 and fluorescein, as well as the simulated IR spectra of the carboxylate anion form and the dianion form of fluorescein (see Figure 2). Fluorescein can exist in seven different forms in dilute aqueous solutions,[30] and which form(s) is/are encapsulated in ZIF-8 could affect the functional performance of the material. In the HCR synthesis employed in this study (see *Experimental Section*), the mIm linker is deprotonated by a stoichiometric amount of triethylamine (Et$_3$N) to reduce reaction time and to improve product yield. The solid product is rigorously washed to remove excess guests. The negative of the logarithm of the acid dissociation constant (pK$_a$) of various forms of fluorescein determines that the carboxylate anion and dianion forms should dominate in aqueous solutions. Since Et$_3$N is a Lewis base and the amount ratio of fluorescein : mIm (or equivalently, fluorescein : Et$_3$N) is very small, it is expected that the majority of fluorescein in the HCR reaction process should be the singly and doubly deprotonated carboxylate anion and dianion forms, respectively.[30] Subsequently, we performed pH response and solvatochromism studies to resolve the component species in the synthesized materials (see Supporting Information). The foregoing chemical arguments form the basis for the selection of fluorescein species involved in the DFT simulations, i.e. the fluorescein species chosen for the calculations are always the carboxylate anion and/or dianion form.

The FTIR spectra feature multiple vibrational modes in the range from 600 cm$^{-1}$ to 1800 cm$^{-1}$. The spectra of the 0.05 mM to 1 mM fluo@ZIF-8 samples are comparable to that of the most diluted 0.01 mM fluo@ZIF-8 sample. However, for the most concentrated sample of 5 mM fluo@ZIF-8, some noticeable differences in its spectrum can be identified (as designated by the asterisks in Figure 2). The most important differences between the spectra of pristine ZIF-8, fluorescein solid sample, 0.01−1 mM and 5 mM fluo@ZIF-8 samples are presented as insets in Figure 2. The simulated IR spectra of the molecular carboxylate anion and dianion forms of fluorescein (at the B3LYP/6-311G$^*$ level of theory, scaled by an empirical factor of 0.97) are also included in the comparison to gain more insight.

First we compare the spectrum for ZIF-8 synthesized using the rapid HCR method to that using the conventional 'slow' method (24 h, without Et$_3$N). The main difference is found at 749 cm$^{-1}$, whose relative peak intensity becomes larger with respect to that for 759 cm$^{-1}$ (see also normalized spectra in Figure S1a). Both modes are ascribed to the out-of-plane ring deformation of the mIm linkages.[36] Besides, the bands at about 994, 1145 and 1310 cm$^{-1}$ become somewhat broader. These observations suggest that there are more framework defects present in ZIF-8 (HCR) than ZIF-8 (slow).[37] However, except for the moderate difference at the 749 cm$^{-1}$ peak, the differences are only slight, indicating that overall the defects are not significant, consistent with the XRD data.

Next we focus on the spectra for ZIF-8 (slow) and fluo@ZIF-8 systems. Consider the relative intensity of the peak at 749 cm$^{-1}$ with respect to that for 759 cm$^{-1}$. ZIF-8 (HCR) and fluo@ZIF-8 samples all exhibit an increase of the former peak intensity, and the difference between different fluo@ZIF-8 samples is marginal, suggesting that the increase of this peak should mainly be ascribed to the formation of more defects in ZIF-8 (HCR), rather than to be associated with the fluorescein guest. At about 1310 cm$^{-1}$, the vibrational modes from ZIF-8 (attributed to in-plane antisymmetric ring stretching)[36] and the anionic and dianionic forms of fluorescein appear to give rise to the somewhat enhanced and broadened peaks at about the same position for the 0.01−1 mM and 5 mM fluo@ZIF-8. At 1180 cm$^{-1}$, the peak is red-shifted with an increasing guest concentration (Figure S1b). A more pronounced effect is observed for the 1400 – 1650 cm$^{-1}$ region, where the two bands (1400 – 1500 cm$^{-1}$ and 1550 – 1650 cm$^{-1}$) are both slightly shifted (also see Figures S1c and S1d). The shifts can be interpreted as a combined effect of the ZIF-8 host framework being perturbed by or interacting with the guest molecules.

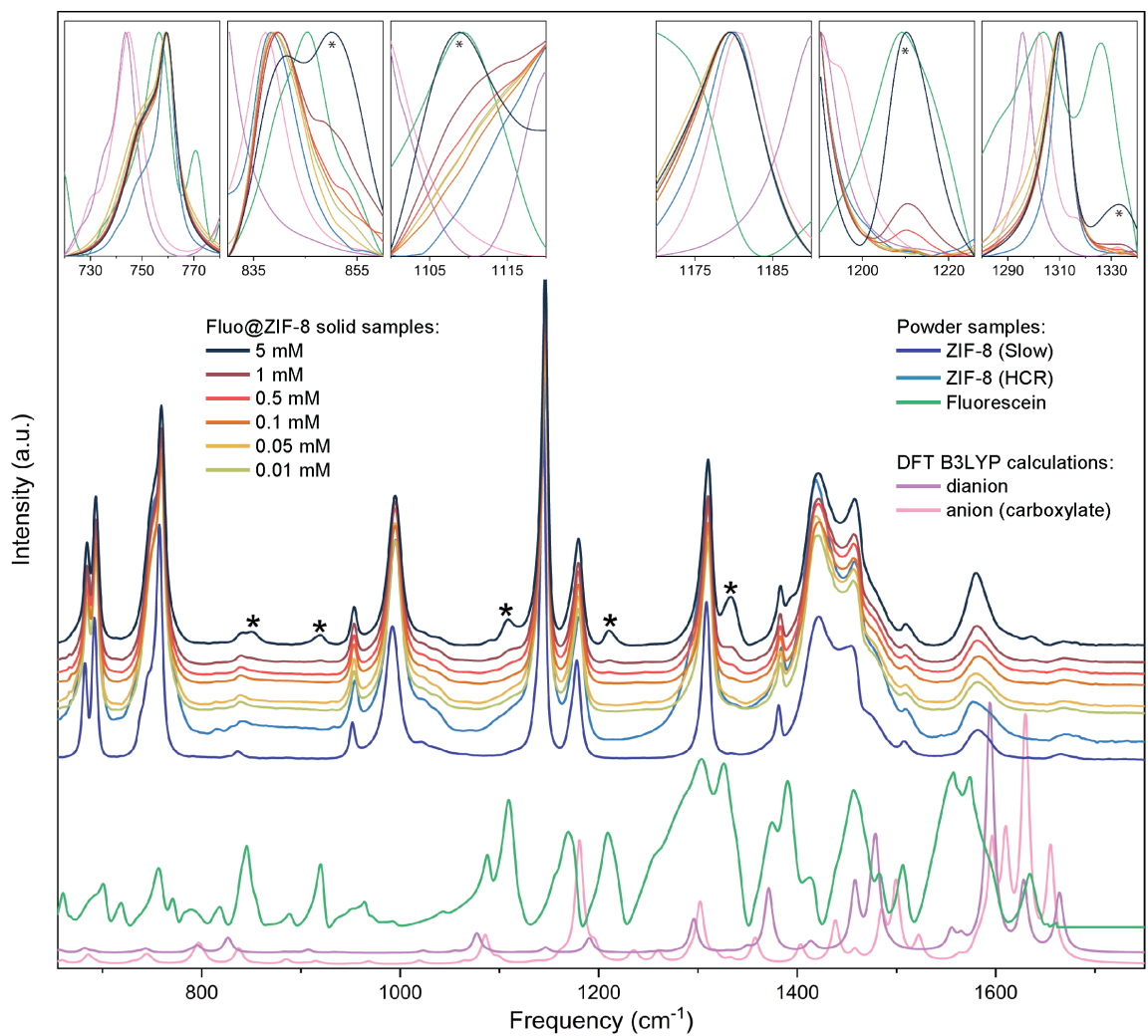

**Figure. 2.** Experimental ATR-FTIR spectra of ZIF-8, fluorescein solid sample, fluo@ZIF-8 synthesized under different conditions, and the simulated IR spectra of the carboxylate anion form and the dianion form of fluorescein. Note the spectra in the insets are normalized. The IR spectra for the carboxylate anion and dianion are computed at the B3LYP/6-311G* level of theory and scaled by an empirical factor of 0.97 for better comparison.

On the other hand, there are a few regions, notably 850, 920, 1110, 1209, and 1333 cm$^{-1}$, where a few more pronounced new features in the spectra of the fluo@ZIF-8 samples evolve. At first glance, one might be inclined to attribute most of these features to the neutral guest fluorescein molecules because, for example, the peaks at 1209 and 1110 cm$^{-1}$ both align almost perfectly. Since the measured fluorescein FTIR spectrum is obtained from a solid sample, it is likely to exhibit some solid-state effect compared to a hypothetical diluted fluorescein sample. The good agreement at these two peaks could be coincidental. However, a few other new peaks, *e.g.* at 850 and 920 cm$^{-1}$, can be visually fitted with the corresponding peaks from the anion or dianion form as well, with somewhat worse quality though. Considering we are directly comparing the simulated spectra for the ionic forms with the measured spectra for the fluorescein, ZIF-8 and fluo@ZIF-8 samples, an imperfect match can be expected. It is worth noting that the attribution of the above new features to fluorescein species is sensible only if there is a positive correlation between the guest loading and the fluorescein amount used in the synthesis, which is supported by the guest loading trend obtained from NMR spectroscopy (Figure 1d).

The comparison in Figure 2 based on molecular DFT calculations above provides motivation for modeling the guest@host system more rigorously. Indeed, the effect of guest-host interactions between the ZIF-8 host and the fluorescein guest can be seen in the periodic DFT simulations (at the PBEsol0-3c level of theory, see the section on *Computational details*) as well (Figure 3). The simulated IR spectra of ZIF-8, fluorescein dianion and a proposed model of fluo@ZIF-8 (*vide infra*), with fluorescein being in the dianion form (together with a $Zn^{2+}$ counter ion) encapsulated in the cage of ZIF-8 are compared with the key peaks/bands highlighted. Note that no scaling factor is used for the simulated spectra presented in Figure 3. A colored region in the IR spectrum of dianion@ZIF-8 corresponds to the region in the spectrum of the dianion or ZIF-8. In particular, several regions show the effect of the contribution by fluorescein, and the corresponding vibrational modes (vectors indicate the atomic vibrations) are illustrated.

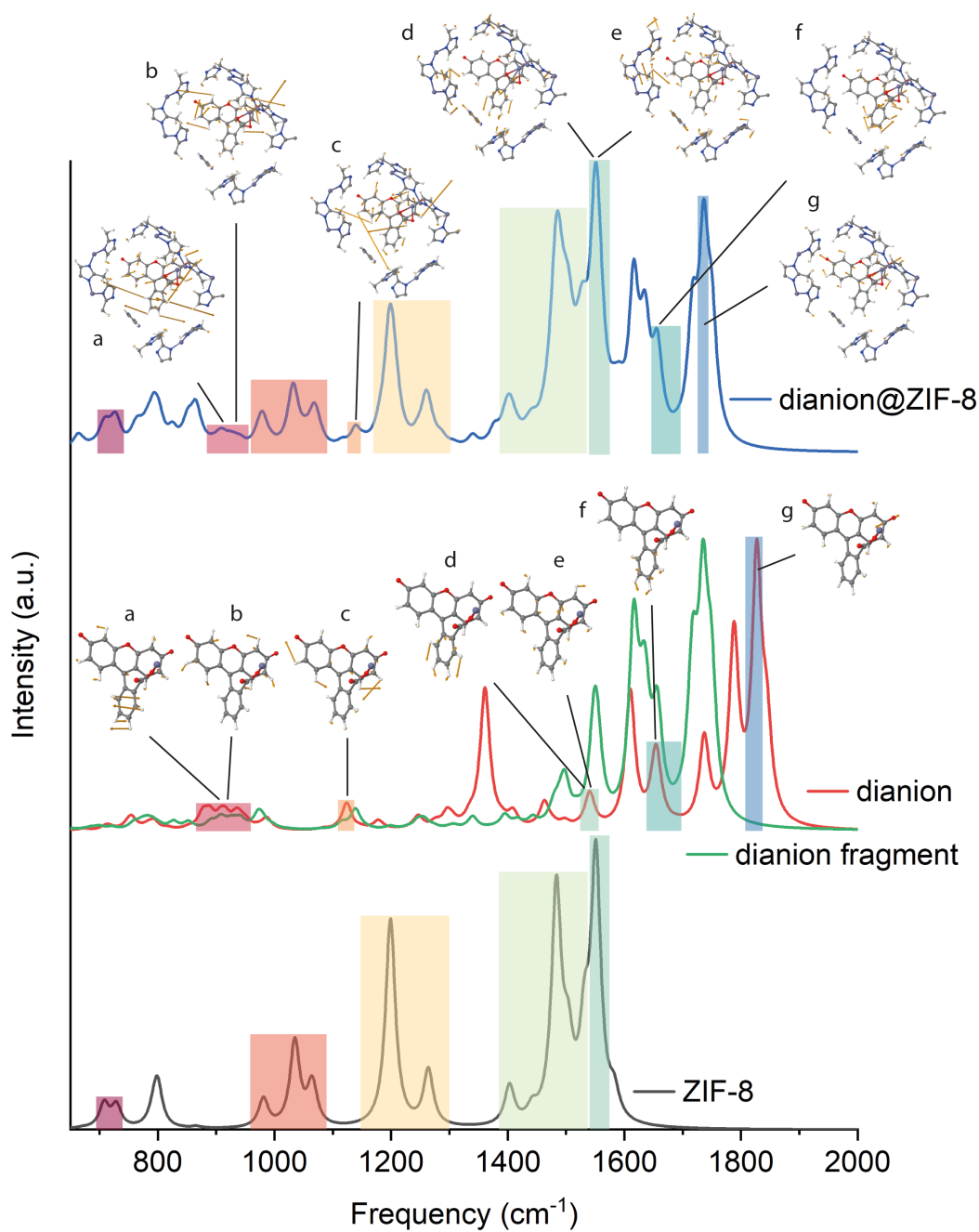

**Figure 3.** Simulated IR spectra of ZIF-8, fluorescein (dianion, with a $Zn^{2+}$ counterion), a model of dianion@ZIF-8 (with the dianionic charge of fluorescein balanced by a $Zn^{2+}$ ion in the same cage), and the fluorescein-$Zn^{2+}$ adduct inside the 'frozen' ZIF-8 framework of the dianion@ZIF-8 model (designated as dianion 'fragment'). All spectra are calculated at the PBEsol0-3c level of theory, and simulated using a Lorentzian profile with a FWHM of 20 cm$^{-1}$. No scaling factor is used for the vibrational frequencies therefore offsets are observed.

The vibrational mode at 909 cm$^{-1}$ for the dianion features an out-of-plane bending mode of the CH groups on the benzenecarboxylate unit in the fluorescein molecule. On the other hand, the mode at 915 cm$^{-1}$ for dianion@ZIF-8 consists of out-of-plane bending of CH groups on both the mIm linker and more importantly the benzenecarboxylate unit. The character of the vibrations due to fluorescein dianion in these two modes (modes a) are similar. Furthermore, the mode at 916 cm$^{-1}$ (mode b) for the dianion can be associated with the mode at 926.5 cm$^{-1}$ (mode b) for dianion@ZIF-8, in which case the contribution from fluorescein dianion is not the out-of-plane bending of the CH groups on the benzenecarboxylate unit, but the in-plane motion of the fused xanthene ring. A connection can thus be made between the regions 870 – 960 cm$^{-1}$ for the dianion and 885 – 950 cm$^{-1}$ for dianion@ZIF-8. Similarly, mode c for the dianion (1124 cm$^{-1}$) is associated with mode c for dianion@ZIF-8 (1139 cm$^{-1}$), as both modes mainly involve the same CH group in-plane bending of the xanthene ring. The highlighted bands containing modes a and b, and the ones containing mode c provide a model to interpret the spectral evolution for the bands at 850 cm$^{-1}$ and 1110 cm$^{-1}$ in Figure 2, respectively. For further insights, we computed the vibrational modes of the fluorescein dianion adduct as in the dianion@ZIF-8 model by decoupling them from the vibrational modes of the framework (designated as dianion fragment, see *Computational details*). The comparison between the dianion molecular adduct and the fluo@ZIF-8 guest fragment shows that the confinement of the adduct in the cage of the ZIF-8 host and the interaction with the host framework lead to a shift of the peaks.

The spectra at the higher frequency region (above 1400 cm$^{-1}$) are more complicated and indicative of a significant overlap of guest and host peaks. The narrow band around 1550 cm$^{-1}$ has its contributing modes from both the ZIF-8 framework and the fluorescein dianion. Two such modes, d and e, are presented along with the spectra. At even higher frequencies (above 1600 cm$^{-1}$), the spectra seem to be dominated by contributions from the fluorescein dianion that are not clearly evident in the experimental spectra because of the dilution of the molecule in the samples. For instance, both modes f (1653 cm$^{-1}$) and g (1736 cm$^{-1}$) for dianion@ZIF-8 have their molecular origin in modes f (1650 cm$^{-1}$) and g (1827 cm$^{-1}$) for the dianion, respectively. However, these two modes for dianion@ZIF-8 somewhat differ from the corresponding modes for the dianion. For instance, vibrations of the right end (from the perspective of Figure 3) of the xanthene ring for modes g are similar, but vibrations of the left end are somewhat enhanced relatively when encapsulated. This can be seen as a manifestation of the interaction between the guest and the host. Roughly speaking, the high frequency region (1400 – 1900 cm$^{-1}$) in the simulated spectra corresponds to the experimentally determined region between 1400 – 1650 cm$^{-1}$ in Figure 2.

It should be noted that a few spectral regions observed in experiments, e.g. the band around 1333 cm$^{-1}$ in Figure 2, still cannot be explained directly, reflecting the possible limitations of the above simplified dianion (with Zn$^{2+}$) and dianion@ZIF-8 (charge balanced by Zn$^{2+}$) models. Or they might be attributed to potential surface aggregates of fluorescein whose IR behavior is likely to resemble that of the fluorescein solid sample. The existence of surface fluorescein for the 5 mM fluo@ZIF-8 is being studied in the next section.

**Photophysical properties of fluorescein@ZIF-8**

The excitation and emission spectra measured from fluorescence spectroscopy are summarized in Figure 4. The excitation peak maximum of fluo@ZIF-8 for all concentrations is at about 499 nm. In terms of the excitation region and the spectral shape, this is in reasonable agreement, considering the difference of the environment of fluorescein, with reported data for fluorescein in dilute aqueous solutions.[30] With an increasing concentration of fluorescein, a shoulder peak at about 465 nm gradually arises, and the excitation spectrum becomes broader at both the left and right hand sides of the maximum; besides, the emission spectra feature a red shift. It is proposed that the shoulder peak is largely due to the H-aggregates of fluorescein, and the broadening at a longer wavelength above 499 nm can be attributed to the J-aggregates, which are responsible for the red shift of the emission peak maximum. This argument will be elaborated shortly afterwards when the lifetime data are discussed. Quantum yield (QY) data in Figure 4d are consistent with the above interpretation because higher concentration of fluorescein favors the formation of aggregates, and the quenching effect should become more prevalent. From the lifetime decay curves in Figure 4e, it can be seen that the fluo@ZIF-8 samples with higher concentration of fluorescein exhibit shorter decay times, indicating a high heterogeneity and the coexistence of different guest species.

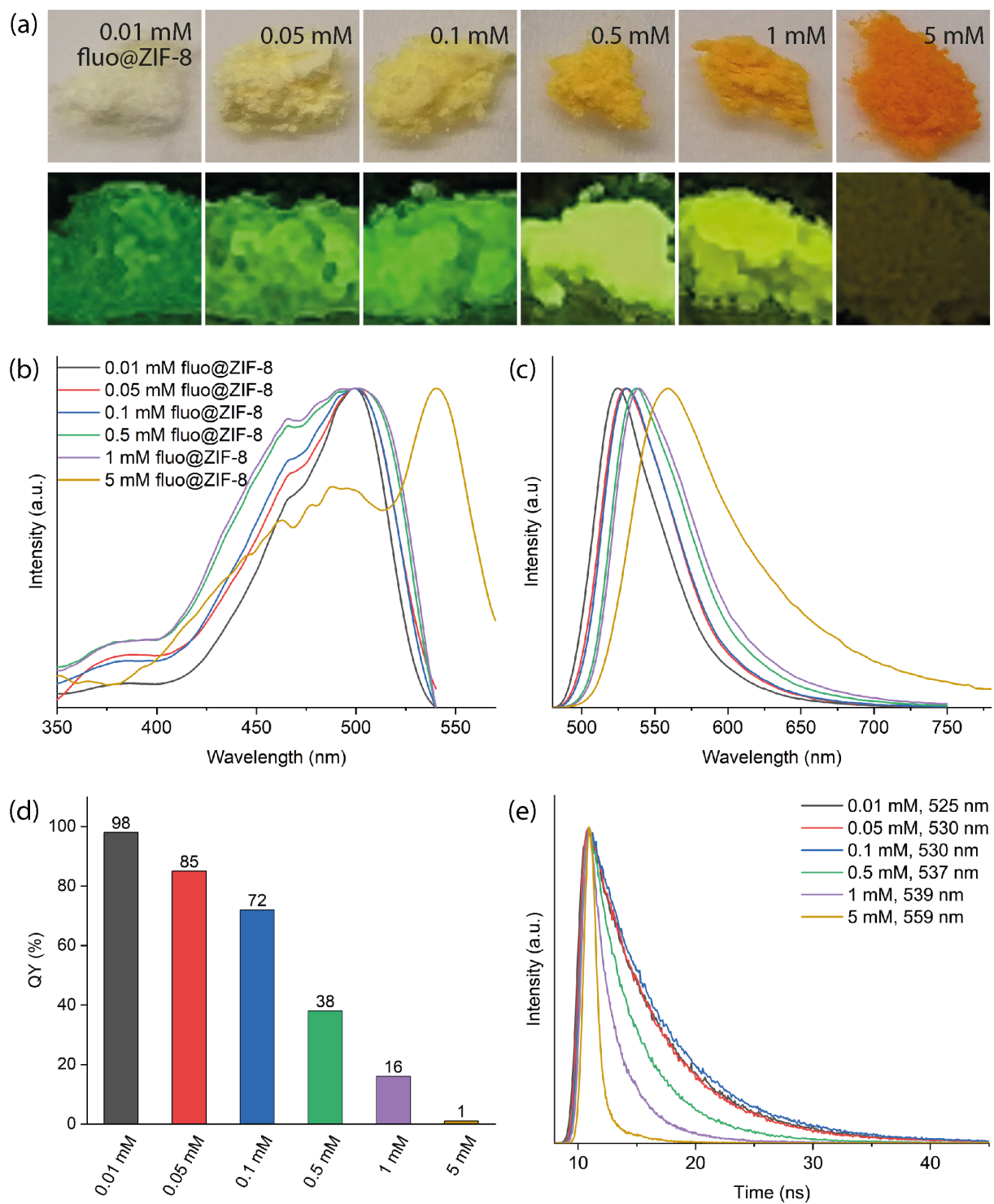

**Figure 4.** (a) Fluo@ZIF-8 samples with different fluorescein guest concentrations (0.01 mM to 5 mM) viewed under natural light (top row) and their luminescence when excited under a 365-nm UV lamp (bottom row). (b) Normalized excitation spectra (observed at emission wavelength of 560 nm, except for 5 mM fluo@ZIF-8, where 580 nm is chosen). (c) Normalized emission spectra (observed at excitation wavelength of 460 nm). (d) Quantum yield of the solid-state samples. (e) Lifetime decay data from TCSPC measurements.

A more detailed analysis of the lifetime data measured from time-correlated single photon counting (TCSPC) are presented in Figure 5. For 0.01 mM fluo@ZIF-8, only two lifetime components, 3.9 ns and 6.5 ns, are observed. The component with lifetime greater than 6 ns persists in all the samples up to 1 mM, while its contribution tends to decrease. From 0.1 mM onwards, a new component, 1.1 ns, appears, which shows an overall trend of increasing contribution up to 1 mM, just like the component above 3 ns. Therefore, it is natural to assign the 6.5 ns component to the fluorescein monomers, whose contribution are decreasing with the concentration due to the formation of aggregates; the 3.9 ns and 1.1 ns time components can be attributed to fluorescein aggregates. Indeed, the 3.9 ns time component should correspond to the deactivation of J-aggregates since those are emissive, while the shorter component is the lifetime of H-aggregates, as they are either non-emissive or only weakly-emissive. The case of 5 mM is somewhat special, as the monomer lifetime is not observed, giving way to aggregates and a new shorter lifetime component of 0.15 ns. We reasoned that this new component must be associated with fluorescein aggregates formed on the outer surface of the fluo@ZIF-8 nanocrystals, since the concentration of fluorescein is so high in this sample that we cannot avoid surface aggregation. It should be noted that we did not rule out the possibility of surface adsorption. However, as the lifetime analysis shows, only when the amount of fluorescein used in the synthesis becomes large does the contribution of surface species become relevant. The photostability study (in a later section) also supports this argument.

Interestingly, since the size of the ZIF-8 pores is relatively smaller than even the smallest dimers of fluorescein (obtained from DFT simulation), the aggregates must be located outside the pores but still contained within the fluo@ZIF-8 system. A possible explanation for this is that fluo@ZIF-8 might form core-shell structures.[38] For instance, it has been demonstrated that even large metal nanoparticles can be entrapped into ZIF-8 crystals.[39,40] Moreover, it is anticipated that the accommodation of aggregates inside the crystal could be facilitated by the formation of framework defects in the composite system of ZIF-8.[36,41]

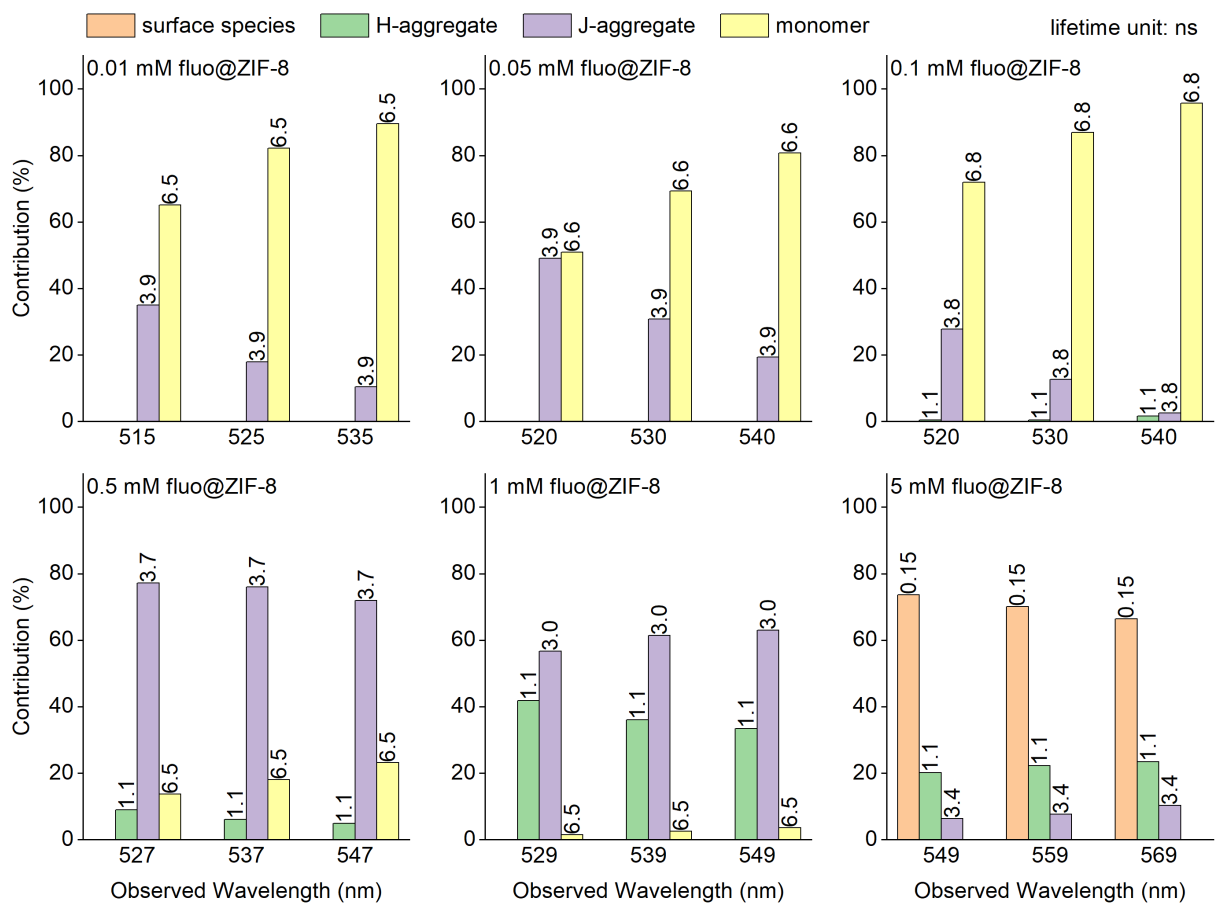

**Figure 5.** Lifetime data representing the contributions from the different forms of fluorescein existing in the fluo@ZIF-8 system.

### Band gaps of fluorescein@ZIF-8

The optical band gaps of the synthesized fluo@ZIF-8 materials are estimated from the Kubelka–Munk (KM) function (Figure 6a). It is worth noting that by increasing the concentration of fluorescein, the band gap of the fluo@ZIF-8 material decreased, from 2.47 eV for 0.01 mM fluo@ZIF-8 to 2.38 eV for 5 mM fluo@ZIF-8. The 'limit' is reached at the pristine fluorescein solid sample, whose band gap is 2.20 eV which is lower than all the above samples. This is in agreement with our previous discussion, since the fluorescein in the guest-encapsulated material should be in closer proximity to each other for higher concentration samples, either because of the higher guest loading of the pores or the formation of more aggregates. The trend of band gap variation with concentration is also consistent with the band gap estimation from the Tauc plot using fluorescein solutions in MeOH (Figure 6b).

To gain further insights, the electronic structure of the dianion@ZIF-8 guest-host model system (see Figure 6c) is computed to compare with the experimental observation. As shown in Figure 6d, the band structure and density of states (DOSs) around the band gap of 3.08 eV are dominated by the electronic levels of the encapsulated molecule (i.e. fluorescein guest). Indeed, the highest-occupied (HOCO) and the lowest-unoccupied (LUCO) crystalline orbitals are clearly localized around fluorescein, as shown in Figure 6e. The band gap is computed at the HSEsol-3c level of theory (Figure 6d in red), which is known to give a better prediction of the band gap in solids. The computed result, 2.52 eV, differs from the experimental one by less than 0.2 eV. Notably, such theoretical findings support the notion that the fluorescein should have been encapsulated in the pores of ZIF-8 by nanoconfinement. Furthermore, supercell calculations reveal that the band gap is not strongly affected by a lower guest loading. For instance, Figure 6f-g show that the results for a 2×1×1 supercell are quite similar to the 1:1 guest-host systems depicted in Figure 6a. Subsequently to simulate a further decrease of the guest loading, we computed for a 2×2×1 and a 2×2×2 supercell that yield a predicted band gap of around 2.5 eV (see Table S4).

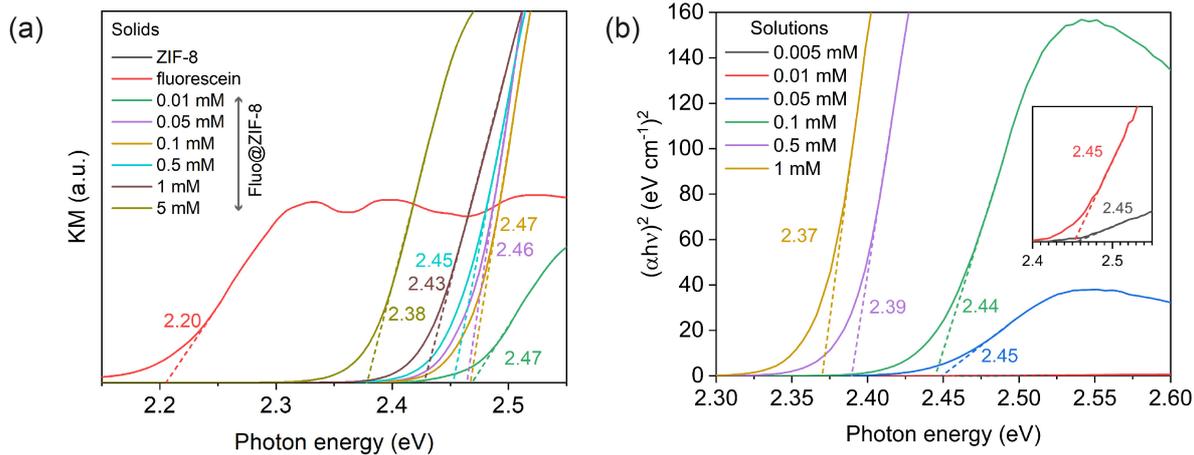
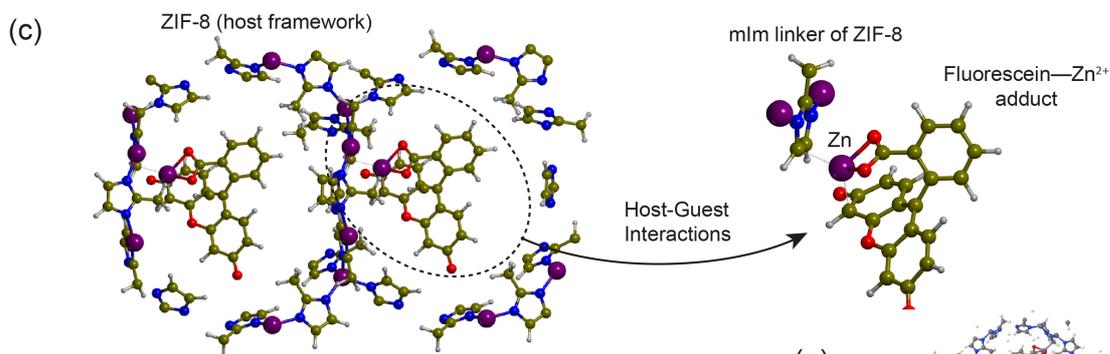
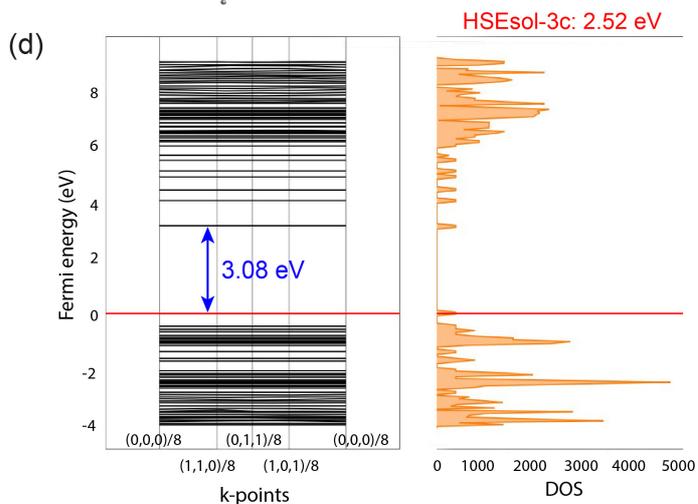
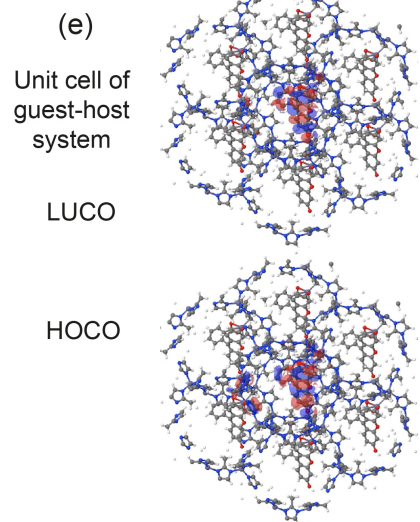
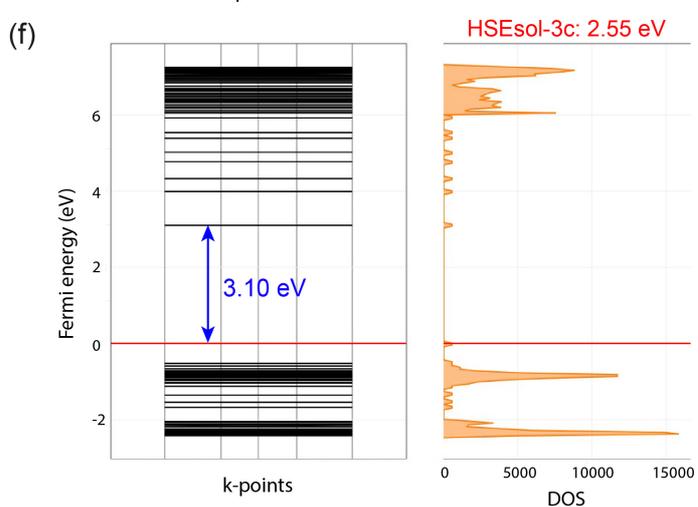
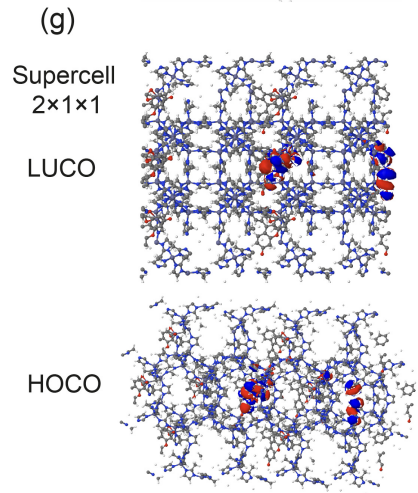

**Figure 6.** (a) Optical band gaps estimated from the Kubelka–Munk (KM) function for fluo@ZIF-8. (b) Optical band gaps estimated from the Tauc plots for fluorescein in MeOH solutions of different concentrations. (c) Structural model for DFT calculations, of fluorescein dianion (with $Zn^{2+}$) encapsulated inside the pore of ZIF-8. Besides the coordination between Zn and the $COO^-$ group of fluorescein, Zn interacts with mIm and the aromatic ring of fluorescein (designated by dashed lines). (d) Band gap of fluorescein (dianion form) encapsulated inside ZIF-8 calculated at the PBEsol0-3c (blue) and HSEsol-3c (red) levels of theory, respectively, with density of states (DOSs). (e) The highest occupied crystalline orbital (HOCO) and the lowest unoccupied crystalline orbital (LUCO), respectively, for (d). (f, g) The same information as in (d) and (e), respectively, for the 2×1×1 supercell of the dianion form of fluorescein encapsulated in ZIF-8.

**Application to solid-state white light emitting devices**

Since fluo@ZIF-8 has remarkably high quantum yield (especially for lower concentration samples, see Figure 4d), this gives us the opportunity to explore the possibility of its application to solid-state WLEDs. The combination of a blue-light LED source and the emission of selected fluo@ZIF-8 thin films (of different concentrations) attached on top of the LED chip gives rise to a multicolor and white-light emitting device (Figure 7a-d). The emission maximum of the blue-source is located at about 449 nm. When it is used to excite the fluo@ZIF-8 film, the emitted light and some of the source light penetrating the film mix and form a combined spectrum as depicted in Figure 7e-f.

By tuning the power of the blue LED (i.e. blue-light source intensity) and the thickness of the fluo@ZIF-8 film, a variety of color temperature in the near region of pure white can be obtained, see Figure 7g. Using thinner photoactive films, a greater amount of blue light (the peak at 449 nm) penetrates the film as shown in Figure 7e. The peaks at larger wavelengths are due to the emission of the fluo@ZIF-8 film. Increasing the power results in a larger relative intensity of the blue-light peak, leading to a point closer to the blue region (the lower left corner) in the CIE diagram. Higher concentration fluo@ZIF-8 samples emit at longer wavelengths. As a result, their corresponding points in the CIE diagram are closer to the red region (the right corner). By tuning the thickness and concentration (emission wavelength) of the sample, a variety of colors can hence be obtained.

When the film of sample is relatively thicker, less blue light passes through the film as can be confirmed in the emission spectra shown in Figure 7f. The resulting color essentially consists of the emission wavelength of the film sample itself. Changing the power of the blue source simply alters the brightness of the emission color, therefore the CIE points of the same thick film largely overlap.

**Photostability study**

The 0.01 mM and 5 mM fluo@ZIF-8 materials are exposed to a 450-nm blue light from the FS5 spectrofluorometer for an extended period of 25 hours with saturated emission signal (see Figure S4). The emission maximum for both samples declines over time, but the emission maximum for 5 mM fluo@ZIF-8 decreases faster initially, in agreement with the previous argument that more surface species of fluorescein should exist (see Figure 5). These surface species should be more vulnerable to photodegradation since they are directly exposed to an environment of oxygen, moisture, etc. In total, both samples have a decrease of the emission maximum by about 10% in this accelerated photodegradation study. Overall, the shielding provided by the nanoconfinement of ZIF-8 framework has considerably improved the photostability of fluorescein compared to the unprotected state (see Figure S4d).

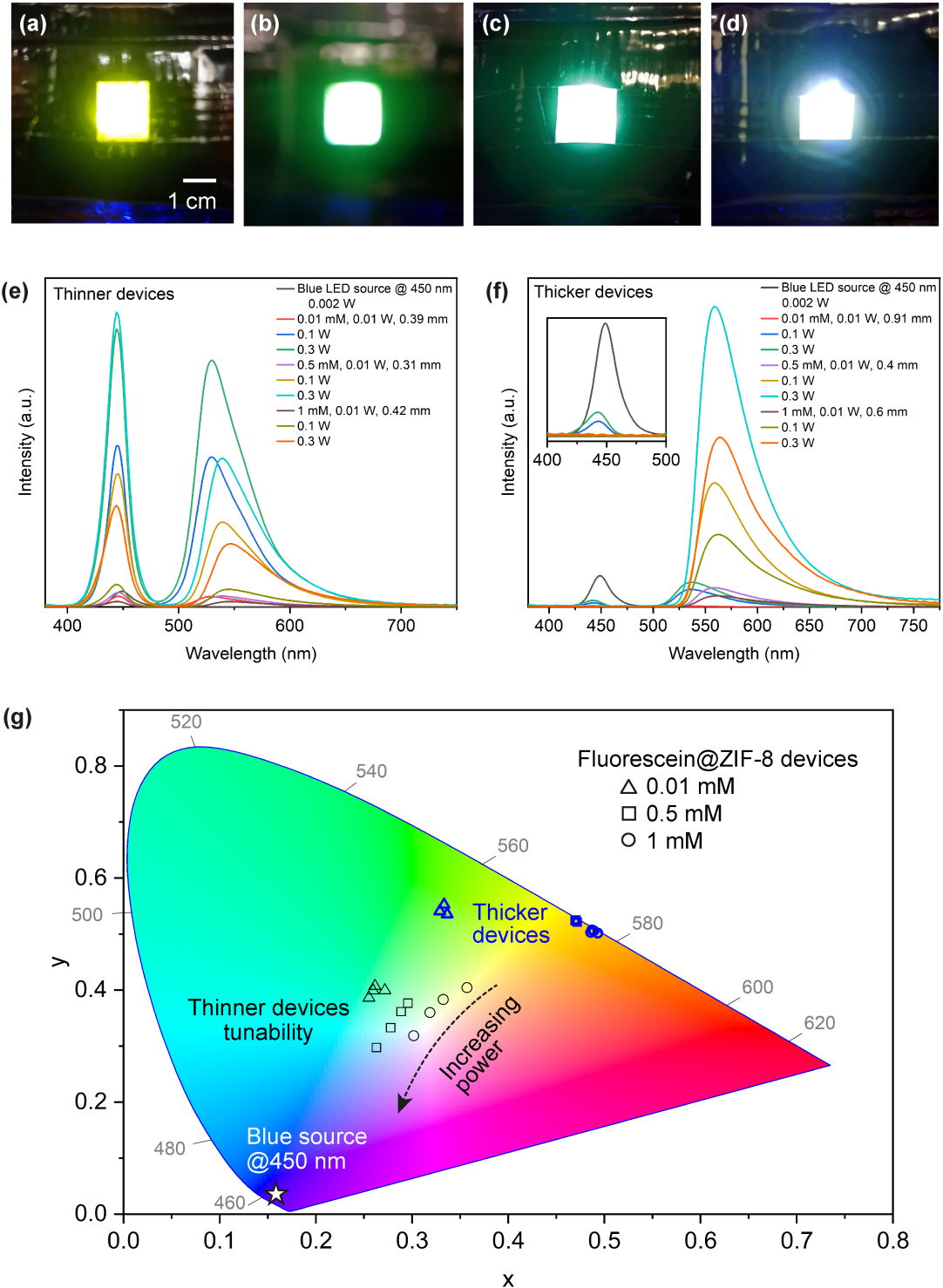

**Figure 7.** (a - d) Prototype solid-state lighting device comprising a thin photoactive film of the selected fluo@ZIF-8 installed in front of a blue-light LED chip, rendering different color temperatures in (g). (e) Emission spectra of a relatively thin film of selected fluo@ZIF-8 of varying concentration excited by a blue 450-nm LED source. (f) Emission spectra similarly obtained using a relatively thick film. (g) CIE 1931 chromaticity diagram of the emission color of the blue LED source coupled with the fluo@ZIF-8 film.

**Conclusions**

A series of dye@MOF composite nanomaterials consisting of fluorescein guests nanoconfined in the ZIF-8 nanocrystals (as a porous host) have been synthesized and systematically studied to understand their photophysical and photochemical properties. The samples have been rigorously washed to minimize the guest residues adhered on the crystal surface. Measured and simulated IR spectra for the relevant components, i.e. fluorescein, ZIF-8, and fluo@ZIF-8, have been compared, and evidence of encapsulation of fluorescein can be found. Electronic excitation and emission measurements, lifetime analysis, pH response and solvatochromism studies of the materials have strengthened this argument. The solid-state nanoparticles could yield a remarkably high QY of ~98% when the fluorescein concentration used in the synthesis is relatively low (0.01 mM). Thin photoactive films of fluo@ZIF-8 and a blue LED chip have been combined to produce a white light with tunable colors. The materials also possess improved photostability compared with the pristine fluorescein due to host shielding. These properties present fluo@ZIF-8 as a promising dye@MOF type tunable phosphor for phosphor-conversion WLEDs and for applications in high-QY photonics.

**Experimental Section**

**Synthesis of fluo@ZIF-8 and ZIF-8**

100 mL of MeOH is mixed with 166.2 mg of fluorescein to produce a stock solution of concentration 5 mM. A series of fluorescein solutions (referred to as solutions A) of different concentrations, namely 1 mM, 0.5 mM, 0.1 mM, 0.05 mM, and 0.01 mM, are then prepared through dilution of the stock solution. 892.5 mg of $Zn(NO_3)_2 \cdot 6H_2O$ is dissolved in 10 mL of MeOH, and 492.6 mg of 2-methylimidazole (mIm) and 0.837 mL of triethylamine ($Et_3N$) are added to 10 mL of MeOH, to form solutions B and C, respectively. Solutions A, B and C are combined in a proportion of 25 mL : 10 mL : 10 mL and stirred for 5 min until the reaction is complete.

The product is centrifuged at 8000 rpm for 5−10 min and the liquid phase is disposed of. The solid product is washed with 40 mL of MeOH, sonicated for 5 min, and centrifuged again under the same conditions. The same procedure is repeated for a total number of 3−5 times until the liquid phase is clear and transparent (for higher concentrations of fluorescein, a few more washing cycles may be needed). After the last centrifugation step, the liquid phase is disposed of and the solid product is dried in an oven at 90 °C for 1 h. The product is cooled to room temperature and gently ground to powder form using a mortar and pestle. The samples are labeled, as 0.01 mM fluo@ZIF-8 for example, according to the concentration of the respective solution A from which they are prepared. The synthesis is summarized in a flowchart in Figure S12.

It is observed that the higher the concentration of solution A, the more washing cycles are needed to remove the remaining fluorescein in the reaction product, which indicates that more fluorescein molecules are adhered on the surface of fluo@ZIF-8 if the sample is prepared from a more concentrated fluorescein solution.

For comparison, pristine ZIF-8 is also synthesized using two methods: the 'rapid' HCR method (with $Et_3N$) and the conventional 'slow' method (without $Et_3N$). The synthesis of ZIF-8 using the HCR method is very similar to that of fluorescein@ZIF-8, except that fluorescein solution (solution A) is not added in the reaction mixture, i.e. only solution B and solution C is combined to form ZIF-8. The slow method to synthesize ZIF-8 is similar to the HCR method, except that no triethylamine ($Et_3N$) is added in solution C, and the reaction time is substantially longer (24 h) compared with the HCR method (5 min).

**Materials characterization**

Powder X-ray diffraction (PXRD) measurements were carried out in a Rigaku MiniFlex X-ray diffractometer with a Cu Kα source (1.541 Å). Fourier-transform infrared (FTIR) spectra were recorded on a Nicolet iS10 FTIR spectrometer equipped with a diamond attenuated total reflectance (ATR) cell. Atomic force microscopy (AFM) micrographs were collected using a Neaspec s-SNOM microscope operating under the tapping mode. The guest loadings were characterized by solution $^1H$ NMR spectroscopy at 298 K using a Bruker Avance NEO spectrometer operating at 600 MHz, equipped with a BBO cryoprobe. Further details on NMR sample preparations and analysis of spectra are given in Supporting Information (Figures S15-S20).

Steady-state fluorescence spectra, steady-state diffuse reflectance spectra, luminescence quantum yield (QY), and time-correlated single photon counting (TCSPC) emission decay data were recorded using the FS-5 spectrofluorometer (Edinburgh Instruments) equipped with the appropriate modules for each specific experiment. For TCSPC measurements, the samples were pumped with a 365 nm EPLED picosecond pulsed laser source. Lifetime fitting of the time constants from decay data was performed using the Fluoracle software.

For MOF-LED device fabrication, a small amount (~mg) of MOF nanoparticles was evenly spread over a thin glass coverslip and covered by another slip to create a thin film 'sandwich' structure. The edges of the coverslips were sealed with a tape to hold the material in place. The thickness of the sandwiched device can be easily adjusted by controlling the amount of nanoparticles used. The sealed device was positioned on top of a 3.6 V blue LED SMD source (450 nm) for optical conversion studies. The produced fluorescence and CIE 1931 were recorded using a UPRtek spectrophotometer (MK350N Plus).

**Computational details**

The molecular structures for the carboxylate anion and the dianion in the gas phase are optimized and subsequent normal mode analysis is performed on the optimized geometries at the B3LYP[42-44]/6-311G*[45,46] level of theory in density functional theory (DFT) using the quantum chemical package Gaussian 16.[47] The IR spectra for the carboxylate anion and dianion in the gas phase are computed and scaled by an empirical factor of 0.97 (Figure 2),[48] and are broadened using a Lorentzian full width at half maximum (FWHM) parameter of 10 cm$^{-1}$. The same level of theory as in the gas phase is used to model the solvation of fluorescein in MeOH and to compute the vertical excitation energies through time-dependent DFT (TDDFT) calculations.

In the solid state, only the p-quinoid and zwitterionic forms of the neutral fluorescein molecule form pure crystalline structures.[49] The lactone form is present only in co-crystals. Therefore, the crystalline structures of the p-quinoid and zwitterionic forms are simulated and the IR spectra calculated and compared with the experimental IR spectrum for the fluorescein solid sample (Figure S9).

The synthesis conditions (presence of the Lewis base - triethylamine (Et$_3$N)) dictate that fluorescein should exist in its dianion and/or carboxylate anion form. A model with a dianion charge balanced by a Zn$^{2+}$ is thus utilized in all relevant simulations for fluo@ZIF-8 and also denoted as dianion@ZIF-8 (Figures 3 and 6). To better elucidate the guest-host interactions in dianion@ZIF-8, simulations of the guest dianion (including a counterion $Zn^{2+}$) in the model dianion@ZIF-8, i.e. the dianion-$Zn^{2+}$ system, are also performed (Figure 3).

In the solid state, DFT calculations are performed at the PBEsol0-3c level of theory.[50] The PBEsol0-3c is a cost-effective solid-state electronic structure method based on the PBEsol0 hybrid functional that includes two semi-classical corrections to take into account dispersive interactions with the D3 scheme[51] and to remove the basis set superposition error (BSSE) through the gCP method.[52] All the calculations are performed with a development version of CRYSTAL17[53] in its massively parallel version, using an all-electron split valence double-zeta basis set recently revised by some of us to deal with solid-state calculations on inorganic and metal-organic systems. For the numerical integration grid of the exchange-correlation part a (75,974) pruned grid is adopted, corresponding to the XLGRID keyword implemented in CRYSTAL. The tolerance for one- and two-electrons integrals is set to $10^{-7}$, $10^{-7}$ for the calculation of Coulomb integrals and to $10^{-7}$, $10^{-7}$ and $10^{-25}$ for the exchange integrals.

A full unconstrained relaxation of both lattice parameters and atomic parameters is performed. Default convergence criteria on gradients and displacements are adopted.

For the calculation of the vibrational frequencies at Γ point the mass-weighted Hessian matrix is computed by numerical differentiation of the analytical first derivatives. The IR intensities are calculated with the Berry phase approach by evaluating the atomic Born tensor as polarization differences between the original and the distorted geometries. IR spectra are simulated using a Lorentzian profile with a FWHM of 20 cm$^{-1}$. For the calculation of the vibrational modes of fluorescein-$Zn^{2+}$ adduct inside the 'frozen' ZIF-8 framework of the dianion@ZIF-8 model (termed dianion 'fragment'), a reduced Hessian matrix is computed for the molecular fragment thus decoupling its vibrations from the framework ones.

Band structure and density of states (DOSs) are computed at the same level of theory to analyze the electronic structure of the examined systems. Computed results are plotted by using CRYSPLOT.[54] JMol[55] is used to plot the crystalline orbitals.

Single-point energy calculations are carried out with the HSEsol-3c method[50] on the optimized model systems. Supercells of different sizes (2×1×1 and 2×2×1) are devised to model the lower content of fluorescein to assess the effect of guest loading on the band gap.

**Supporting information**

FTIR spectra, excitation-emission spectra, lifetime constants, pH response, CIE chromaticity, solvatochromic studies, photostability data, additional DFT simulations and analysis, TGA, flow chart of synthesis route, additional DFT calculations, DFT input/output files, NMR spectra and analysis.

**Conflicts of Interest**

There are no conflicts of interest to declare.

**Acknowledgements**

This work is supported by the ERC Consolidator Grant through the grant agreement 771575 (PROMOFS). The authors would like to acknowledge the use of the University of Oxford Advanced Research Computing (ARC) facility in carrying out this work (http://dx.doi.org/10.5281/zenodo.22558).

**References**

(1) Nakamura, S.; Senoh, M.; Iwasa, N.; Nagahama, S.-i.; Yamada, T.; Mukai, T. Superbright green InGaN single-quantum-well-structure light-emitting diodes. *Jpn. J. Appl. Phys.* **1995**, *34*, L1332-L1335.

(2) Ye, S.; Xiao, F.; Pan, Y. X.; Ma, Y. Y.; Zhang, Q. Y. Phosphors in phosphor-converted white light-emitting diodes: Recent advances in materials, techniques and properties. *Mater. Sci. Eng. R Rep* **2010**, *71*, 1-34.

(3) Sakuma, K.; Omichi, K.; Kimura, N.; Ohashi, M.; Tanaka, D.; Hirosaki, N.; Yamamoto, Y.; Xie, R.-J.; Suehiro, T. Warm-white light-emitting diode with yellowish orange SiAlON ceramic phosphor. *Opt. Lett.* **2004**, *29*, 2001-2003.

(4) Chen, L.-Y.; Chang, J.-K.; Cheng, W.-C.; Huang, J.-C.; Huang, Y.-C.; Cheng, W.-H. Chromaticity tailorable glass-based phosphor-converted white light-emitting diodes with high color rendering index. *Opt. Express* **2015**, *23*, A1024-A1029.

(5) Balaram, V. Rare earth elements: A review of applications, occurrence, exploration, analysis, recycling, and environmental impact. *Geosci. Front.* **2019**, *10*, 1285-1303.

(6) Ogi, T.; Kaihatsu, Y.; Iskandar, F.; Wang, W.-N.; Okuyama, K. Facile synthesis of new full-color-emitting BCNO phosphors with high quantum efficiency. *Adv. Mater.* **2008**, *20*, 3235-3238.

(7) Etchart, I.; Hernandez, I.; Huignard, A.; Berard, M.; Gillin, W. P.; Curry, R. J.; Cheetham, A. K. Efficient oxide phosphors for light upconversion; green emission from $Yb^{3+}$ and $Ho^{3+}$ co-doped Ln(2)BaZnO(5) (Ln = Y, Gd). *J. Mater. Chem.* **2011**, *21*, 1387-1394.

(8) Lin, C.; Yu, M.; Cheng, Z.; Zhang, C.; Meng, Q.; Lin, J. Bluish-white emission from radical carbonyl impurities in amorphous $Al_2O_3$ prepared via the pechini-type sol−gel process. *Inorg. Chem.* **2008**, *47*, 49-55.

(9) Chouhan, N.; Lin, C. C.; Hu, S.-F.; Liu, R.-S. A rare earth-free GaZnON phosphor prepared by combustion for white light-emitting diodes. *J. Mater. Chem. B* **2015**, *3*, 1473-1479.

(10) Zhang, X.; Lu, Z.; Liu, H.; Lin, J.; Xu, X.; Meng, F.; Zhao, J.; Tang, C. Blue emitting BCNO phosphors with high quantum yields. *J. Mater. Chem. B* **2015**, *3*, 3311-3317.

(11) Nyalosaso, J. L.; Boonsin, R.; Vialat, P.; Boyer, D.; Chadeyron, G.; Mahiou, R.; Leroux, F. Towards rare-earth-free white light-emitting diode devices based on the combination of dicyanomethylene and pyranine as organic dyes supported on zinc single-layered hydroxide. *Beilstein J. Nanotechnol.* **2019**, *10*, 760-770.

(12) Wang, Z.; Wang, Z.; Lin, B.; Hu, X.; Wei, Y.; Zhang, C.; An, B.; Wang, C.; Lin, W. Warm-white-light-emitting diode based on a dye-loaded metal-organic framework for fast white-light communication. *ACS Appl. Mater. Interfaces* **2017**, *9*, 35253-35259.

(13) Chaudhari, A. K.; Tan, J. C. Dual‐guest functionalized zeolitic imidazolate framework‐8 for 3D printing white light‐emitting composites. *Adv. Opt. Mater.* **2020**, *8*, 1901912.

(14) Gutiérrez, M.; Martín, C.; Van der Auweraer, M.; Hofkens, J.; Tan, J.-C. Electroluminescent guest@MOF nanoparticles for thin film optoelectronics and solid-state lighting. *Adv. Opt. Mater.* **2020**, *8*, 2000670.

(15) Wang, Z.; Zhu, C. Y.; Mo, J. T.; Fu, P. Y.; Zhao, Y. W.; Yin, S. Y.; Jiang, J. J.; Pan, M.; Su, C. Y. White-light emission from dual-way photon energy conversion in a dye-encapsulated metal-organic framework. *Angew. Chem. Int. Ed.* **2019,** *58*, 9752-9757.

(16) Wen, Y.; Sheng, T.; Zhu, X.; Zhuo, C.; Su, S.; Li, H.; Hu, S.; Zhu, Q. L.; Wu, X. Introduction of red-green-blue fluorescent dyes into a metal-organic framework for tunable white light emission. *Adv. Mater.* **2017,** *29*, 1700778.

(17) Gong, Q.; Hu, Z.; Deibert, B. J.; Emge, T. J.; Teat, S. J.; Banerjee, D.; Mussman, B.; Rudd, N. D.; Li, J. Solution processable MOF yellow phosphor with exceptionally high quantum efficiency. *J. Am. Chem. Soc.* **2014,** *136*, 16724-16727.

(18) Zhang, Y.; Gutiérrez, M.; Chaudhari, A. K.; Tan, J.-C. Dye-encapsulated zeolitic imidazolate framework (ZIF-71) for fluorochromic sensing of pressure, temperature, and volatile solvents. *ACS Appl. Mater. Interfaces* **2020,** *12*, 37477-37488.

(19) Hu, M.-L.; Razavi, S. A. A.; Piroozzadeh, M.; Morsali, A. Sensing organic analytes by metal–organic frameworks: A new way of considering the topic. *Inorg. Chem. Front.* **2020,** *7*, 1598-1632.

(20) Karmakar, A.; Samanta, P.; Dutta, S.; Ghosh, S. K. Fluorescent "turn-on" sensing based on metal-organic frameworks (MOFs). *Chem.-Asian J.* **2019,** *14*, 4506-4519.

(21) Babal, A. S.; Souza, B. E.; Möslein, A. F.; Gutiérrez, M.; Frogley, M. D.; Tan, J.-C. Broadband dielectric behavior of an MIL-100 metal–organic framework as a function of structural amorphization. *ACS Appl. Electron. Mater.* **2021,** *3*, 1191-1198.

(22) Stassen, I.; Burtch, N.; Talin, A.; Falcaro, P.; Allendorf, M.; Ameloot, R. An updated roadmap for the integration of metal-organic frameworks with electronic devices and chemical sensors. *Chem. Soc. Rev.* **2017,** *46*, 3185-3241.

(23) Pascanu, V.; González Miera, G.; Inge, A. K.; Martín-Matute, B. Metal–organic frameworks as catalysts for organic synthesis: A critical perspective. *J. Am. Chem. Soc.* **2019,** *141*, 7223-7234.

(24) Titov, K.; Eremin, D. B.; Kashin, A. S.; Boada, R.; Souza, B. E.; Kelley, C. S.; Frogley, M. D.; Cinque, G.; Gianolio, D.; Cibin, G.; Rudic, S.; Ananikov, V. P.; Tan, J.-C. OX-1 metal-organic framework nanosheets as robust hosts for highly active catalytic palladium species. *ACS Sustain. Chem. Eng.* **2019,** *7*, 5875-5885.

(25) Ma, S.; Zhou, H.-C. Gas storage in porous metal–organic frameworks for clean energy applications. *Chem. Commun.* **2010,** *46*, 44-53.

(26) Zhuang, J.; Kuo, C.-H.; Chou, L.-Y.; Liu, D.-Y.; Weerapana, E.; Tsung, C.-K. Optimized metal–organic-framework nanospheres for drug delivery: Evaluation of small-molecule encapsulation. *ACS Nano* **2014,** *8*, 2812-2819.

(27) Souza, B. E.; Donà, L.; Titov, K.; Bruzzese, P.; Zeng, Z.; Zhang, Y.; Babal, A. S.; Möslein, A. F.; Frogley, M. D.; Wolna, M.; Cinque, G.; Civalleri, B.; Tan, J.-C. Elucidating the drug release from metal–organic framework nanocomposites via in situ synchrotron microspectroscopy and theoretical modeling. *ACS Appl. Mater. Interfaces* **2020,** *12*, 5147-5156.

(28) Chaudhari, A. K.; Ryder, M. R.; Tan, J. C. Photonic hybrid crystals constructed from in situ host-guest nanoconfinement of a light-emitting complex in metal-organic framework pores. *Nanoscale* **2016,** *8*, 6851-6859.

(29) Allendorf, M. D.; Foster, M. E.; Leonard, F.; Stavila, V.; Feng, P. L.; Doty, F. P.; Leong, K.; Ma, E. Y.; Johnston, S. R.; Talin, A. A. Guest-induced emergent properties in metal-organic frameworks. *J. Phys. Chem. Lett.* **2015,** *6*, 1182-1195.

(30) Sjöback, R.; Nygren, J.; Kubista, M. Absorption and fluorescence properties of fluorescein. *Spectrochim. Acta A* **1995,** *51*, L7-L21.

(31) Klonis, N.; Sawyer, W. H. Spectral properties of the prototropic forms of fluorescein in aqueous solutions. *J. Fluoresc.* **1996,** *6*, 147-157.

(32) Togashi, D. M.; Szczupak, B.; Ryder, A. G.; Calvet, A.; O'Loughlin, M. Investigating tryptophan quenching of fluorescein fluorescence under protolytic equilibrium. *J. Phys. Chem. A* **2009,** *113*, 2757-2767.

(33) Chaudhari, A. K.; Kim, H. J.; Han, I.; Tan, J. C. Optochemically responsive 2D nanosheets of a 3D metal-organic framework material. *Adv. Mater.* **2017,** *29*, 1701463.

(34) Chaudhari, A. K.; Han, I.; Tan, J. C. Multifunctional supramolecular hybrid materials constructed from hierarchical self-ordering of in situ generated metal-organic framework (MOF) nanoparticles. *Adv. Mater.* **2015,** *27*, 4438-4446.

(35) Holder, C. F.; Schaak, R. E. Tutorial on powder x-ray diffraction for characterizing nanoscale materials. *ACS Nano* **2019,** *13*, 7359-7365.

(36) Möslein, A. F.; Gutiérrez, M.; Cohen, B.; Tan, J.-C. Near-field infrared nanospectroscopy reveals guest confinement in metal–organic framework single crystals. *Nano Lett.* **2020,** *20*, 7446-7454.

(37) Hu, Y.; Kazemian, H.; Rohani, S.; Huang, Y. N.; Song, Y. In situ high pressure study of ZIF-8 by ftir spectroscopy. *Chem. Commun.* **2011,** *47*, 12694-12696.

(38) El-Toni, A. M.; Habila, M. A.; Labis, J. P.; Alothman, Z. A.; Alhoshan, M.; Elzatahry, A. A.; Zhang, F. Design, synthesis and applications of core–shell, hollow core, and nanorattle multifunctional nanostructures. *Nanoscale* **2016,** *8*, 2510-2531.

(39) Lu, G.; Li, S.; Guo, Z.; Farha, O. K.; Hauser, B. G.; Qi, X.; Wang, Y.; Wang, X.; Han, S.; Liu, X.; DuChene, J. S.; Zhang, H.; Zhang, Q.; Chen, X.; Ma, J.; Loo, S. C.; Wei, W. D.; Yang, Y.; Hupp, J. T.; Huo, F. Imparting functionality to a metal-organic framework material by controlled nanoparticle encapsulation. *Nat. Chem.* **2012,** *4*, 310-316.

(40) Zheng, G.; de Marchi, S.; Lopez-Puente, V.; Sentosun, K.; Polavarapu, L.; Perez-Juste, I.; Hill, E. H.; Bals, S.; Liz-Marzan, L. M.; Pastoriza-Santos, I.; Perez-Juste, J. Encapsulation of single plasmonic nanoparticles within ZIF-8 and SERS analysis of the MOF flexibility. *Small* **2016,** *12*, 3935-3943.

(41) Zhang, C. Y.; Han, C.; Sholl, D. S.; Schmidt, J. R. Computational characterization of defects in metal-organic frameworks: Spontaneous and water-induced point defects in ZIF-8. *J. Phys. Chem. Lett.* **2016,** *7*, 459-464.

(42) Becke, A. D. Density-functional exchange-energy approximation with correct asymptotic behavior. *Phys. Rev. A* **1988,** *38*, 3098-3100.

(43) Vosko, S. H.; Wilk, L.; Nusair, M. Accurate spin-dependent electron liquid correlation energies for local spin density calculations: A critical analysis. *Can. J. Phys.* **1980,** *58*, 1200-1211.

(44) Lee, C.; Yang, W.; Parr, R. G. Development of the colle-salvetti correlation-energy formula into a functional of the electron density. *Phys. Rev. B* **1988,** *37*, 785-789.

(45) McLean, A. D.; Chandler, G. S. Contracted gaussian basis sets for molecular calculations. I. Second row atoms, z=11–18. *J. Chem. Phys.* **1980,** *72*, 5639-5648.

(46) Krishnan, R.; Binkley, J. S.; Seeger, R.; Pople, J. A. Self‑consistent molecular orbital methods. XX. A basis set for correlated wave functions. *J. Chem. Phys.* **1980,** *72*, 650-654.

(47) Frisch, M. J.; Trucks, G. W.; Schlegel, H. B.; Scuseria, G. E.; Robb, M. A.; Cheeseman, J. R.; Scalmani, G.; Barone, V.; Petersson, G. A.; Nakatsuji, H.; Li, X.; Caricato, M.; Marenich, A. V.; Bloino, J.; Janesko, B. G.; Gomperts, R.; Mennucci, B.; Hratchian, H. P.; Ortiz, J. V.; Izmaylov, A. F.; Sonnenberg, J. L.; Williams; Ding, F.; Lipparini, F.; Egidi, F.; Goings, J.; Peng, B.; Petrone, A.; Henderson, T.; Ranasinghe, D.; Zakrzewski, V. G.; Gao, J.; Rega, N.; Zheng, G.; Liang, W.; Hada, M.; Ehara, M.; Toyota, K.; Fukuda, R.; Hasegawa, J.; Ishida, M.; Nakajima, T.; Honda, Y.; Kitao, O.; Nakai, H.; Vreven, T.; Throssell, K.; Montgomery Jr., J. A.; Peralta, J. E.; Ogliaro, F.; Bearpark, M. J.; Heyd, J. J.; Brothers, E. N.; Kudin, K. N.; Staroverov, V. N.; Keith, T. A.; Kobayashi, R.; Normand, J.; Raghavachari, K.; Rendell, A. P.; Burant, J. C.; Iyengar, S. S.; Tomasi, J.; Cossi, M.; Millam, J. M.; Klene, M.; Adamo, C.; Cammi, R.; Ochterski, J. W.; Martin, R. L.; Morokuma, K.; Farkas, O.; Foresman, J. B.; Fox, D. J. *Gaussian 16 rev. C.01*, Wallingford, CT, **2016,**

(48) Rauhut, G.; Pulay, P. Transferable scaling factors for density functional derived vibrational force fields. *J. Phys. Chem.* **1995,** *99*, 3093-3100.

(49) Arhangelskis, M.; Eddleston, M. D.; Reid, D. G.; Day, G. M.; Bucar, D. K.; Morris, A. J.; Jones, W. Rationalization of the color properties of fluorescein in the solid state: A combined computational and experimental study. *Chem.* **2016,** *22*, 10065-10073.

(50) Doná, L.; Brandenburg, J. G.; Civalleri, B. Extending and assessing composite electronic structure methods to the solid state. *J. Chem. Phys.* **2019,** *151*, 121101.

(51) Grimme, S.; Antony, J.; Ehrlich, S.; Krieg, H. A consistent and accurate ab initio parametrization of density functional dispersion correction (DFT-D) for the 94 elements H-Pu. *J. Chem. Phys.* **2010,** *132*, 154104.

(52) Kruse, H.; Grimme, S. A geometrical correction for the inter- and intra-molecular basis set superposition error in hartree-fock and density functional theory calculations for large systems. *J. Chem. Phys.* **2012,** *136*, 154101.

(53) Dovesi, R.; Erba, A.; Orlando, R.; Zicovich-Wilson, C. M.; Civalleri, B.; Maschio, L.; Rérat, M.; Casassa, S.; Baima, J.; Salustro, S.; Kirtman, B. Quantum-mechanical condensed matter simulations with CRYSTAL. *WIREs Comput. Mol. Sci.* **2018,** *8*, e1360.

(54) Beata, G.; Perego, G.; Civalleri, B. Crysplot: A new tool to visualize physical and chemical properties of molecules, polymers, surfaces, and crystalline solids. *J. Comput. Chem.* **2019,** *40*, 2329-2338.

(55) Hanson, R. M.; Lu, X.-J. Dssr-enhanced visualization of nucleic acid structures in Jmol. *Nucleic Acids Res.* **2017,** *45*, W528-W533.

*Supporting Information*

*for*

# Tunable Fluorescein-Encapsulated Zeolitic Imidazolate Framework-8 Nanoparticles for Solid-State Lighting

Tao Xiong,[a,†] Yang Zhang,[a,†] Lorenzo Donà,[b] Mario Gutiérrez,[a] Annika F. Möslein,[a]

Arun S. Babal,[a] Nader Amin,[c] Bartolomeo Civalleri,[b] and Jin-Chong Tan[*,a]

[a]Multifunctional Materials & Composites (MMC) Laboratory, Department of Engineering Science, University of Oxford, Parks Road, Oxford OX1 3PJ, U.K.

[b]Department of Chemistry, NIS and INSTM Reference Centre, University of Turin, via Pietro Giuria 7, Torino 10125, Italy.

[c]Department of Chemistry, University of Oxford, Mansfield Road, Oxford OX1 3TA, U.K.

[†]These authors contributed equally to this work

[*]jin-chong.tan@eng.ox.ac.uk

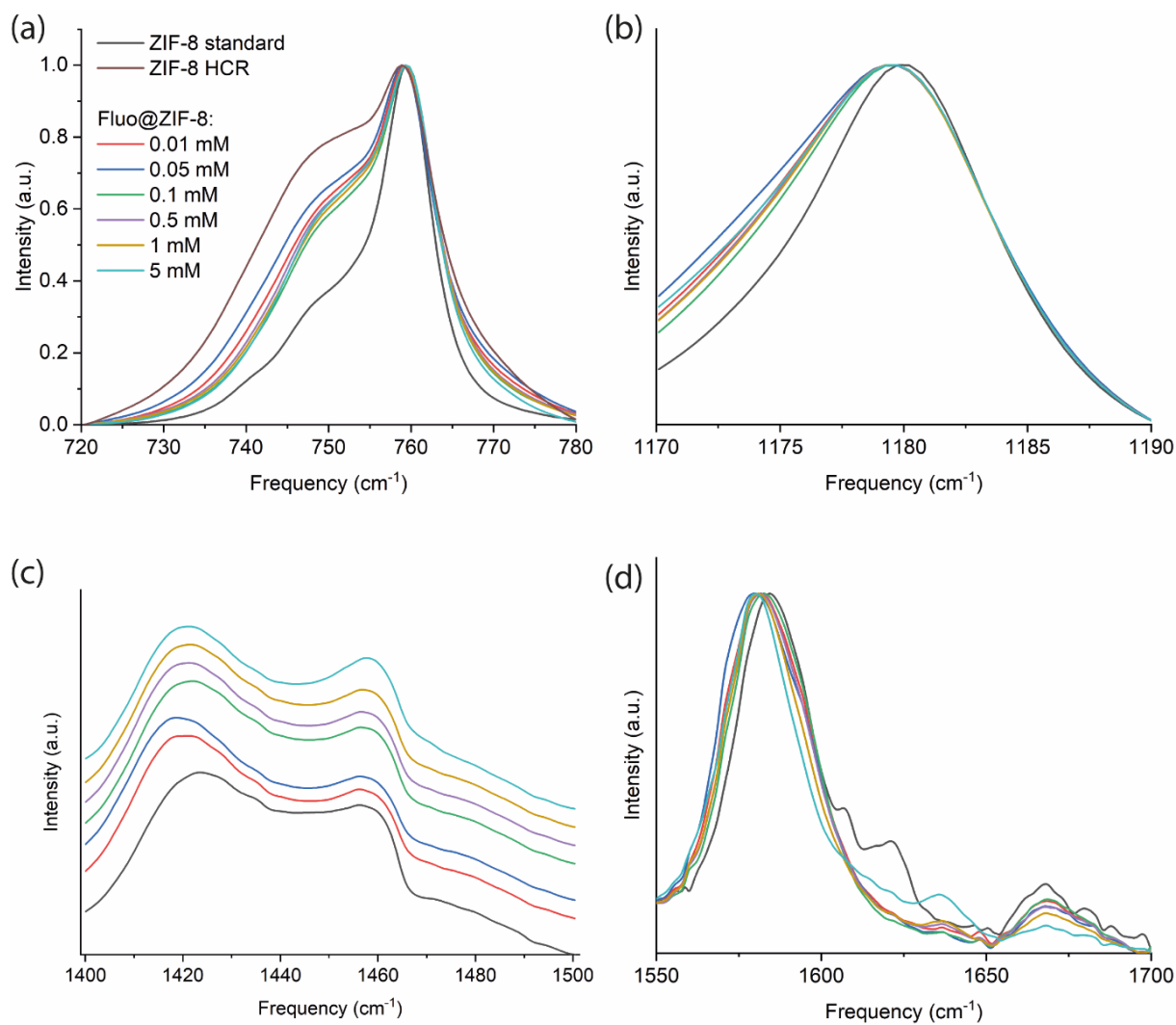

Figure S1. Effects of inclusion of fluorescein guest on the IR spectrum of fluo@ZIF-8. (a) Change of the 749 cm$^{-1}$ : 759 cm$^{-1}$ peak intensity ratio. (a) Red shift of the 1180 cm$^{-1}$ peak. (b) Red and blue shifts of the 1420 cm$^{-1}$ and 1460 cm$^{-1}$ peaks, respectively. (c) Red shift of the 1584 cm$^{-1}$ peak. Note that the 'standard' ZIF-8 sample was obtained using the conventional 'slow' synthesis of 24 hours and without the use of Et$_3$N.

Table S1. Values of lifetime constants ($\tau_i$) and fractional contributions ($a_i$) of the corresponding emission decay components of the fluo@ZIF-8 samples upon excitation at 362.5 nm, obtained from a multi-exponential fitting function, $I(t) = \sum_i a_i e^{-t/\tau_i}$, where $I(t)$ is the photon counts.

| Fluo@ZIF-8 (Fluorescein concentration, mM) | λ/nm | $\tau_1$/ns | $a_1$/% | $\tau_2$/ns | $a_2$/% | $\tau_3$/ns | $a_3$/% |
|---|---|---|---|---|---|---|---|
| 0.01 mM | 515 | | | 3.9 | 34.97 | 6.5 | 65.03 |
| | 525 | | | 3.9 | 17.86 | 6.5 | 82.14 |
| | 535 | | | 3.9 | 10.45 | 6.5 | 89.55 |
| 0.05 mM | 520 | | | 3.9 | 49.05 | 6.6 | 50.95 |
| | 530 | | | 3.9 | 30.77 | 6.6 | 69.23 |
| | 540 | | | 3.9 | 19.31 | 6.6 | 80.69 |
| 0.1 mM | 520 | 1.1 | 0.40 | 3.8 | 27.70 | 6.8 | 71.90 |
| | 530 | 1.1 | 0.47 | 3.8 | 12.60 | 6.8 | 86.93 |
| | 540 | 1.1 | 1.65 | 3.8 | 2.58 | 6.8 | 95.77 |
| 0.5 mM | 527 | 1.1 | 9.03 | 3.7 | 77.24 | 6.5 | 13.73 |
| | 537 | 1.1 | 6.02 | 3.7 | 75.98 | 6.5 | 18.00 |
| | 547 | 1.1 | 4.91 | 3.7 | 71.98 | 6.5 | 23.11 |
| 1 mM | 529 | 1.1 | 41.79 | 3.0 | 56.73 | 6.5 | 1.48 |
| | 539 | 1.1 | 36.06 | 3.0 | 61.47 | 6.5 | 2.47 |
| | 549 | 1.1 | 33.46 | 3.0 | 62.99 | 6.5 | 3.55 |
| 5 mM | 549 | 0.15 | 73.63 | 1.1 | 20.08 | 3.4 | 6.28 |
| | 559 | 0.15 | 70.12 | 1.1 | 22.29 | 3.4 | 7.60 |
| | 569 | 0.15 | 66.39 | 1.1 | 23.37 | 3.4 | 10.24 |

**Methods for pH response and solvatochromic studies**

5 mg of 0.01 mM fluo@ZIF-8 is added to 20 mL of McIlvaine buffer solutions[1] with pH 4, 6, 7, 8, and 10, respectively. The mixture is sonicated before use for excitation and emission measurements.

The sample preparation step for solvatochromism study is very similar, except that buffer solutions are replaced by a variety of solvents, respectively, to form a suspension of the sample after sonication. More specifically, 5 mg of 0.01 mM fluo@ZIF-8 is added to 20 mL of a variety of solvents, respectively, to form the sample suspension.

The excitation and emission spectra of the suspensions in buffer solutions and in solvents, respectively, are measured and compared to that of the solid sample of 0.01 mM fluo@ZIF-8.

**pH response of fluorescein@ZIF-8**

The spectral response of the 0.01 mM fluo@ZIF-8 sample to pH is investigated (Figure S2). From neutral to basic conditions (pH = 7, 8, 10), the excitation spectra show no significant changes with respect to each other. This reflects the fact that the dianion form is favored at more basic condition. It has been reported that when the pH is below 6.43, the anion form starts to dominate until the pH is below 4.31 when the neutral forms contribute more[2]. At pH = 6, the excitation spectrum already shows a higher contribution at larger excitation energies. At pH = 4, the framework structure of ZIF-8 eventually decomposes[3] and the encapsulated fluorescein (in whatever form) is released in the aqueous solution and the spectrum agrees with reported data. At all pH values considered in the experiment, the excitation spectrum is blue-shifted by 0.1 – 0.2 eV. This could be a solvent effect (see the next section on solvatochromism), since water is a polar protic solvent, just like MeOH.

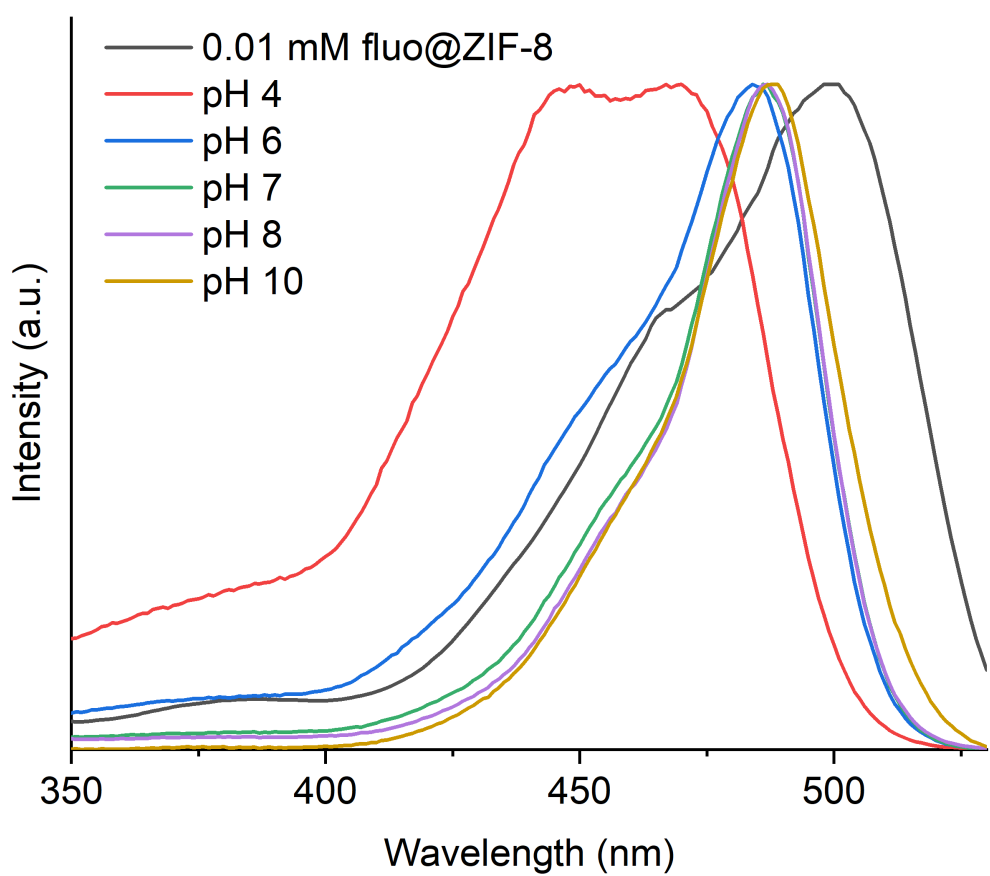

Figure S2. The normalized excitation spectra (observed at emission wavelength 560 nm) of 0.01 mM fluo@ZIF-8 and its suspension in buffer solutions with pH values of 4, 6, 7, 8 and 10, respectively.

**Solvatochromism of fluorescein@ZIF-8**

The effect of a variety of solvents on the excitation and emission spectra of the lowest concentration sample 0.01 mM fluo@ZIF-8 is studied (Figure S3). It turns out that the excitation wavelength of the suspension of 0.01 mM fluo@ZIF-8 in MeOH is blue-shifted by 0.02 – 0.05 eV for each solvent, with the largest shift occurring for MeOH (9 nm, 0.05 eV). The emission spectra undergo a similar blue shift by 0.05 – 0.06 eV.

An explicit solvation model is utilized to understand this observation (Tables S2-S3). On the one hand, one MeOH solvent molecule is introduced to interact with different functional groups of the carboxylate anion and dianion, respectively. On the other hand, five MeOH solvent molecules are included in the model and they can interact with different functional groups of the fluorescein species simultaneously. Using time-dependent DFT (TDDFT), the vertical excitation energy to the lowest bright state (using a criterion of oscillator strength f ≥ 0.1) of the isolated fluorescein species (anion or dianion) is compared with each of its solvation model. Generally, while the excitation energy for the solvated dianion is similar to the isolated one, it is increased by up to 0.19 eV for the solvated anion. Combined with the evidence and argument presented in the main manuscript, it can be expected that the MeOH solvent molecules affect the excitation spectrum of fluo@ZIF-8 initially *via* interaction through the window apertures (~3.4 Å) of the sodalite cage. Although the kinetic diameter of the MeOH molecule (~3.6 Å) is relatively larger than allowed by the windows of ZIF-8 to enter the pores, methanol could still penetrate the narrow apertures of ZIF-8 over time due to lattice dynamics of the flexible framework enabled by gate-opening mechanism.[4,5] Furthermore, the simulated excitation energy shift exhibits the same trend as the experimental observation. The blue shift (0.02 – 0.05 eV) for the latter is smaller than the solvation model, which is expected because (the experimental) interaction through the window is less effective than (the simulated) direct interaction.

The emission spectra also support the above explanation. The emission wavelength for each polar protic solvent is blue-shifted (due to the fluorescein-solvent interaction) but the extent of shift is not too different (because only the -OH group of the solvent molecule is needed to interact with the fluorescein inside the pores so that the size of the solvent molecule plays a relatively minor role). The CIE chart of the suspension (Figure S3d) shows that although for polar aprotic solvents the pattern is less regular, the color rendering performance of the polar protic solvents are more similar as evidenced from the clustering of their coordinates, consistent with the similar nature of the solvent molecular structure.

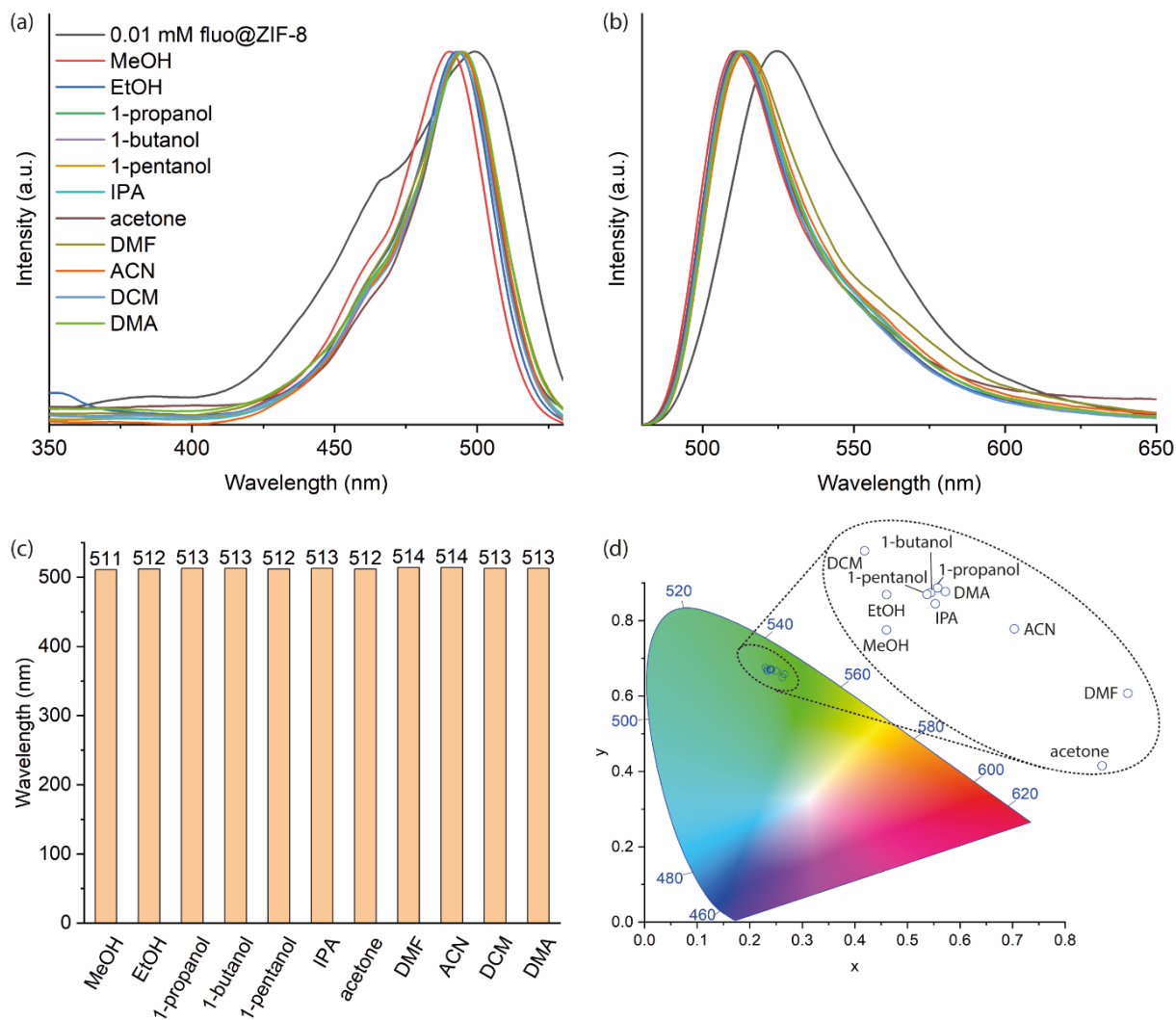

Figure S3. (a) The excitation spectra (observed at emission wavelength 560 nm) of 0.01 mM fluo@ZIF-8 suspension in a variety of solvents. (b) The corresponding emission spectra (excited at 460 nm). (c) The emission maxima in the respective solvents. (d) The CIE 1931 chromaticity diagram of the suspensions.

Table S2. The ground state geometry of fluorescein carboxylate anion in the gas phase (A) and solvated by one MeOH molecule (A1-A3) and five MeOH molecules (A4). The vertical excitation energy $\Delta E_\text{vert}$ refers to the vertical excitation at the respective configuration to the lowest bright excited state with oscillator strength ≥ 0.1. TDDFT calculations performed at the B3LYP/6-311G* level of theory.

| Label | Configuration | $\Delta E_\text{vert}$ (eV) | $\Delta E_\text{vert}(\text{solvation}) - \Delta E_\text{vert}(\text{gas})$ (eV) |
|---|---|---|---|
| A | 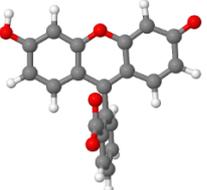 | 2.8732 | NA |
| A1 | 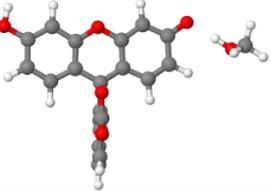 | 3.0061 | 0.1329 |
| A2 | 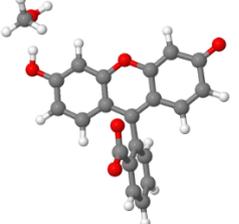 | 2.8735 | 0.0003 |
| A3 | 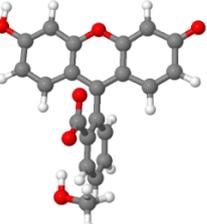 | 2.9433 | 0.0701 |
| A4 | 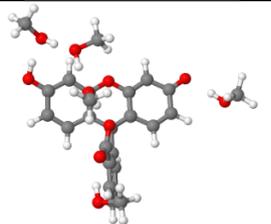 | 3.0671 | 0.1939 |

Table S3. DFT calculated ground state geometry of fluorescein dianion in the gas phase (D) and solvated by one MeOH molecule (D1, D2) and five MeOH molecules (D3). The vertical excitation energy $\Delta E_{vert}$ refers to the vertical excitation at the respective configuration to the lowest bright excited state with oscillator strength $\geq 0.1$. TDDFT calculations performed at the B3LYP/6-311G* level of theory.

| Label | Configuration | $\Delta E_{vert}$ (eV) | $\Delta E_{vert}(\text{solvation}) - \Delta E_{vert}(\text{gas})$ (eV) |
|---|---|---|---|
| D | 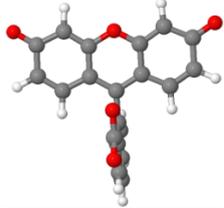 | 2.9461 | NA |
| D1 | 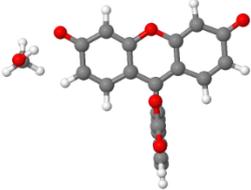 | 2.9400 | -0.0061 |
| D2 | 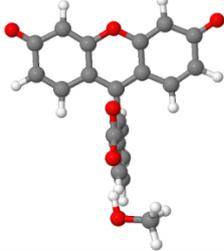 | 2.9561 | 0.0100 |
| D3 | 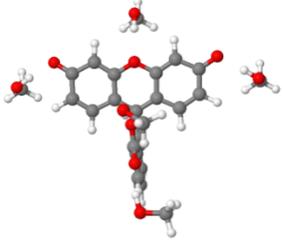 | 2.9485 | 0.0024 |

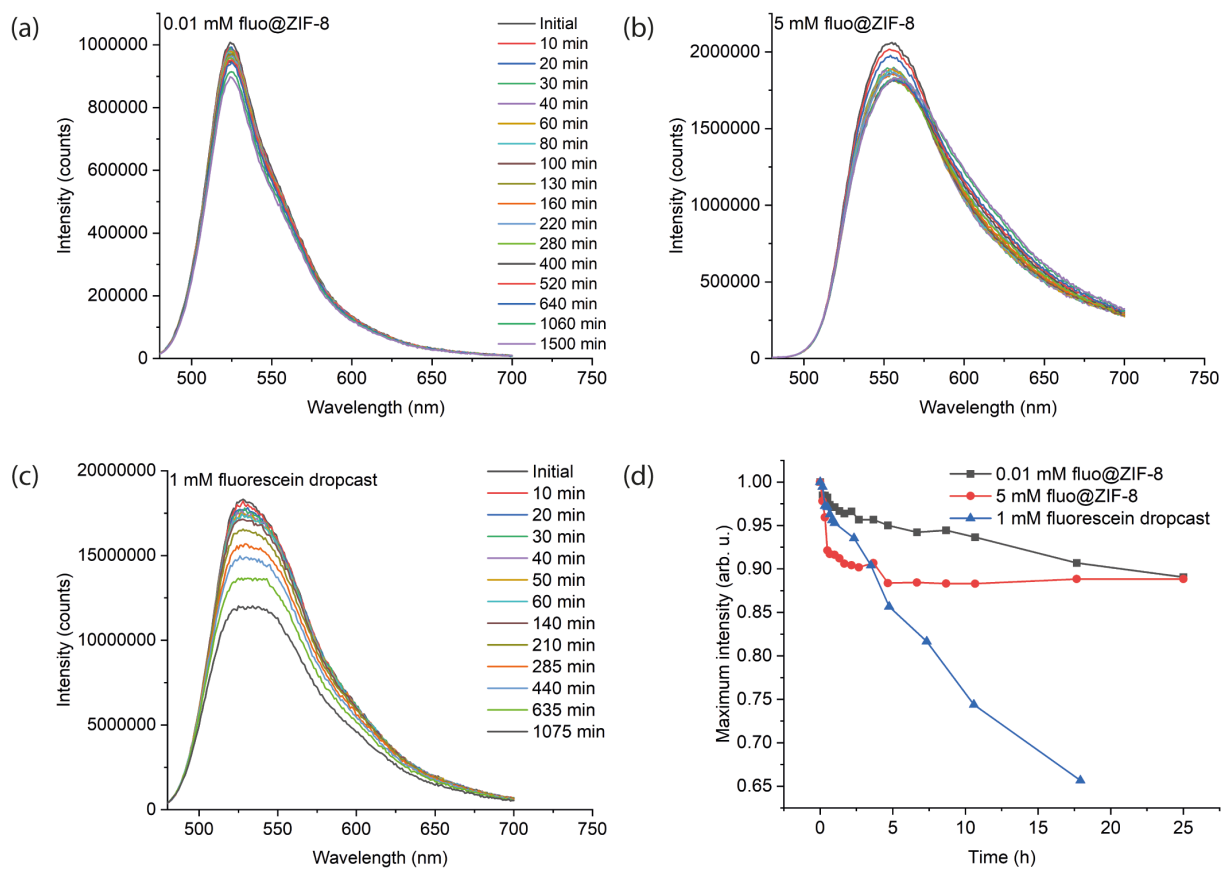

Figure S4. Photostability studies showing the emission spectra of the samples under 450 nm excitation wavelength measured during a prolonged UV exposure time of up to 25 hours. (a) 0.01 mM fluo@ZIF-8. (b) 5 mM fluo@ZIF-8. (c) 1 mM fluorescein in MeOH solution drop casted onto a filter paper. (d) UV-induced photodegradation as a function of exposure time under a Xenon lamp, showing the normalized maximum emission intensity initially set to 1 million counts per second. The lines are guides for the eye.

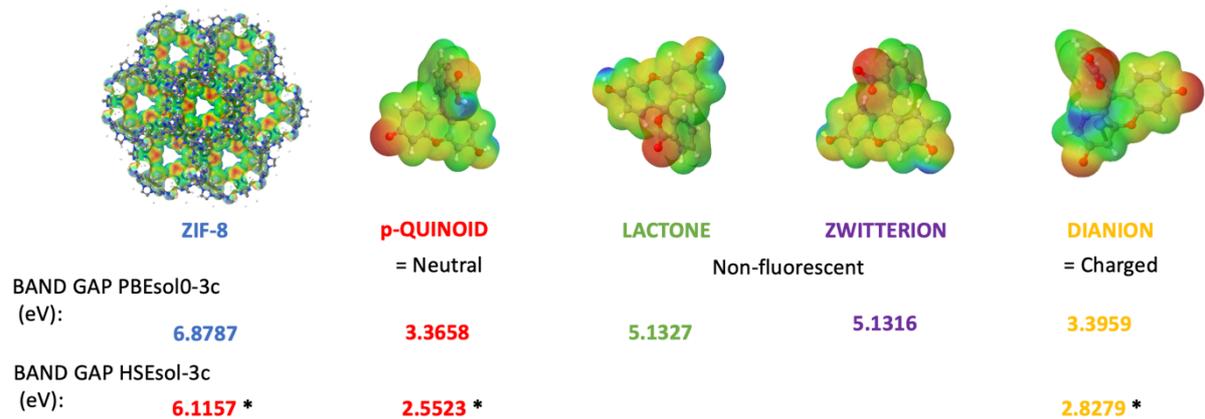

Figure S5. Electrostatic Potential Map (EPM) calculated from CRYSTAL17 code for the ZIF-8 host and the different possible forms of the fluorescein guest. *No geometry optimization performed, band gap obtained from single point energy calculation.

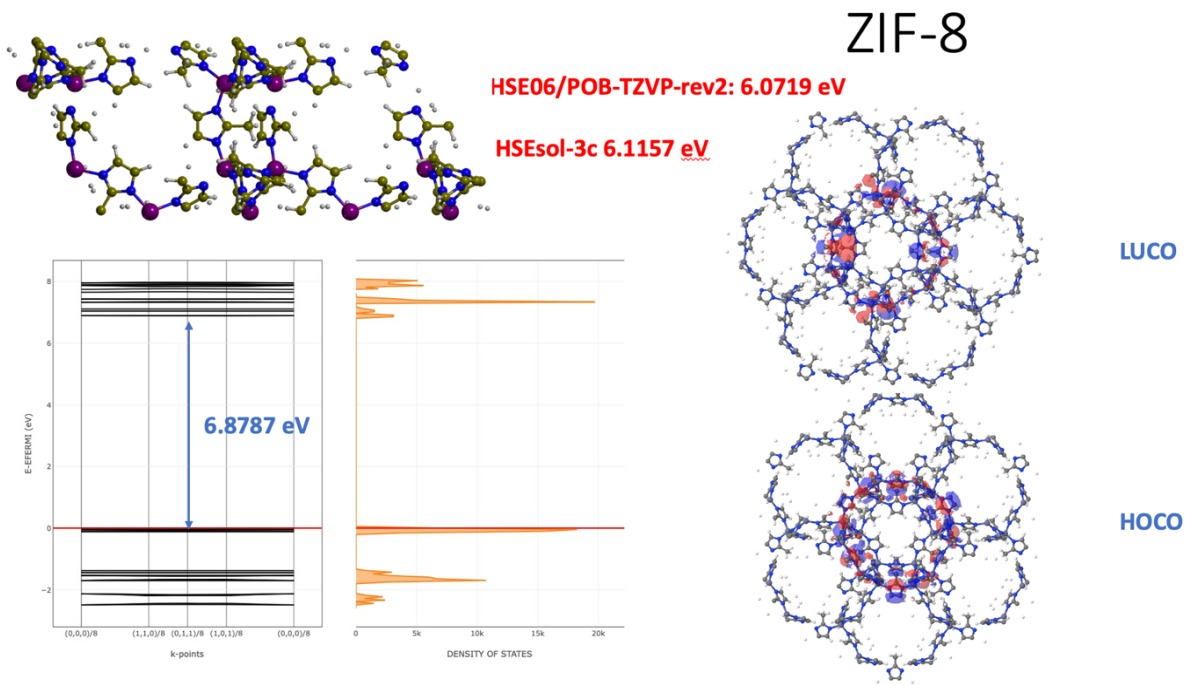

Figure S6. Theoretical calculations of bandgap and LUCO-HOCO crystal orbitals of the porous ZIF-8 framework (host). All periodic DFT calculations employed dispersion interaction using the CRYSTAL17 code. Note the crystal orbitals are localized to the 2-methylimidazolate (mIm) linkers.

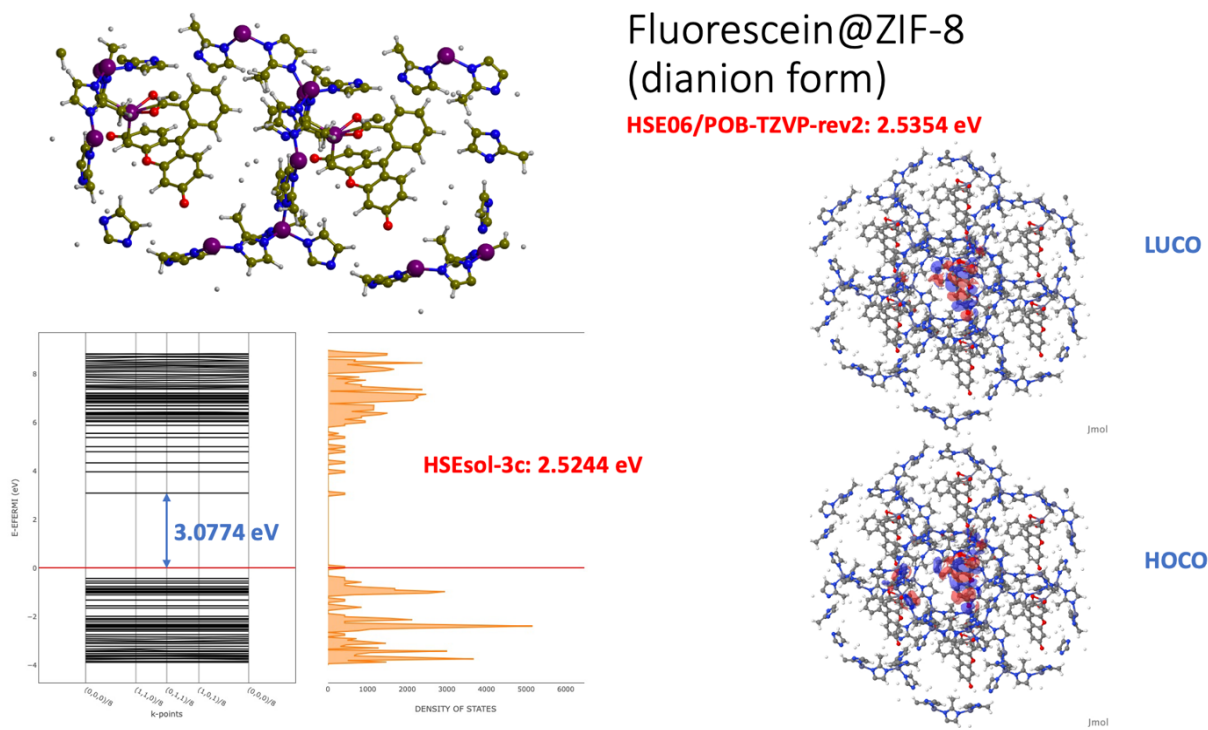

Figure S7. Theoretical DFT calculations of bandgap and LUCO-HOCO crystal orbitals of the fluo@ZIF-8 system. All periodic DFT calculations employed dispersion interaction using the CRYSTAL17 code. Note the crystal orbitals are localized on the fluorescein guest confined in the pore of ZIF-8.

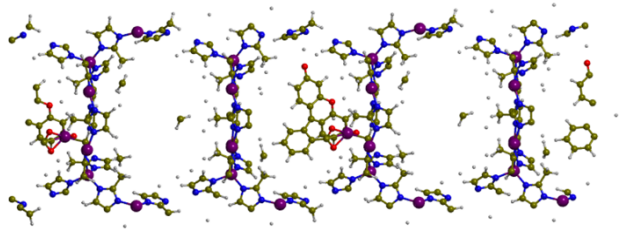
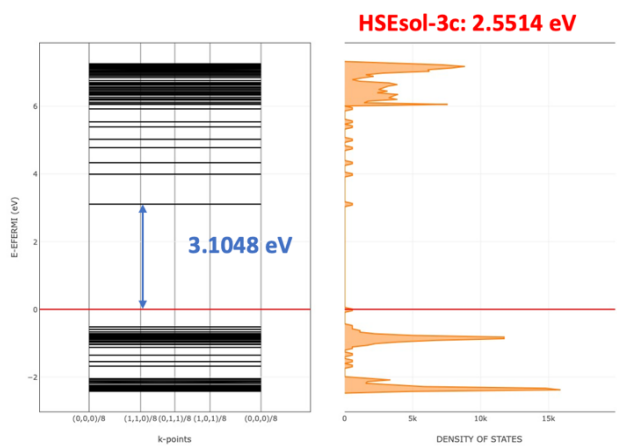
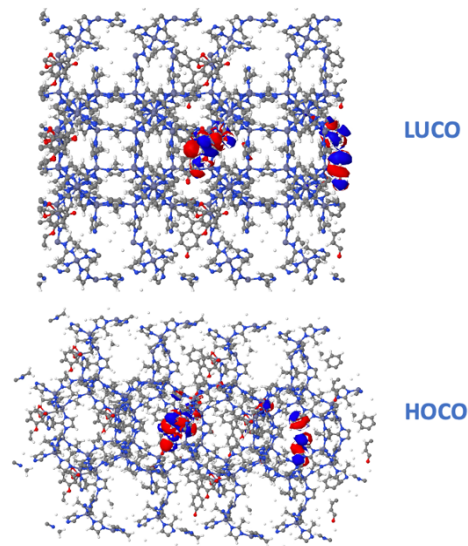

Figure S8. Theoretical DFT calculations of bandgap and LUCO-HOCO crystal orbitals of the 2×1×1 supercell of the fluo@ZIF-8 system. The crystal orbitals are localized on the fluorescein guest molecules confined in the pores of ZIF-8. All periodic DFT calculations employed dispersion interaction using the CRYSTAL17 code.

Table S4. Band gap (in eV) from periodic DFT calculations of the fluorescein dianion@ZIF-8 model system for different loading of the guest molecule confined within the unit cell (of the ZIF-8 host).

| Supercell | 1×1×1 | 2×1×1 | 2×2×1 [a] | 2×2×2 [a] |
|---|---|---|---|---|
| **Guest loading** | 100% | 50% | 25% | 12.5% |
| PBEsol0-3c | 3.077 | 3.105 | 3.093 | 3.090 |
| HSEsol-3c [b] | 2.524 | 2.551 | 2.539 | 2.536 |

[a] The geometry of the dianion@ZIF–8 has been fixed at one of the 2×1×1 supercell (see Figure S8)

[b] Single-point energy calculations on the PBEsol0-3c optimized geometries.

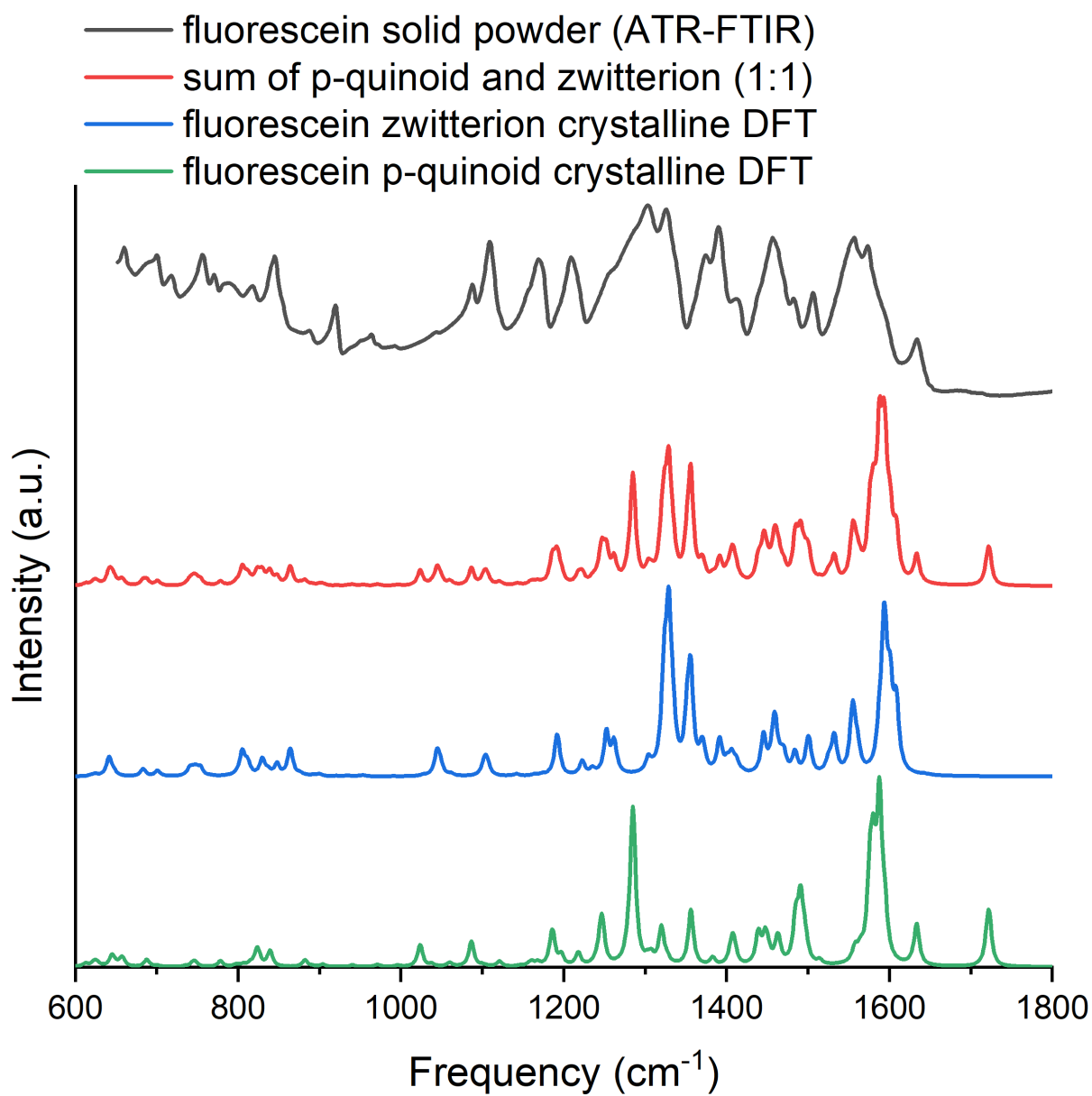

Figure S9. Comparison of the IR spectra of fluorescein solid sample (measured by ATR-FTIR experiment), p-quinoid, zwitterion, and the 1:1 sum of p-quinoid and zwitterion crystalline molecular forms (DFT simulations at PBEsol0-3c level of theory). The simulated spectra have been arbitrarily scaled by a factor of 0.945 for better comparison. A Lorentzian with a FWHM of 10 cm$^{-1}$ is used to broaden the spectra.

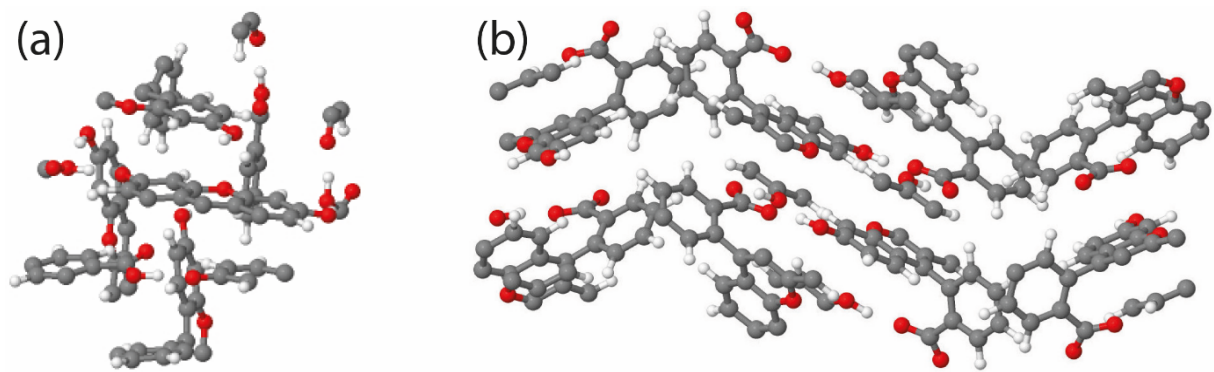

Figure S10. Optimized crystalline structure obtained at the PBEsol0-3c level of theory of fluorescein (a) p-quinoid, and (b) zwitterion.

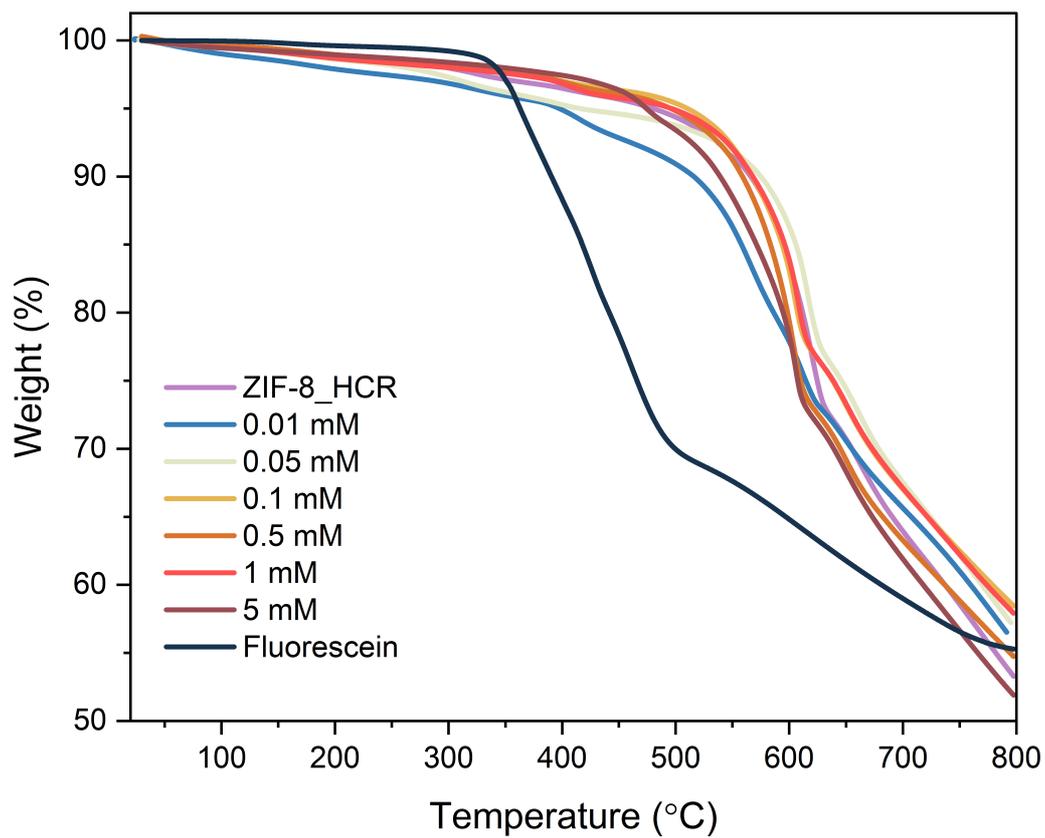

Figure S11. TGA for pristine ZIF-8 (HCR), fluorescein powder, and fluo@ZIF-8 samples. Note the gentle weight loss of ZIF-8 below ~300 °C can be attributed to the loss of occluded solvent molecules and Et$_3$N.

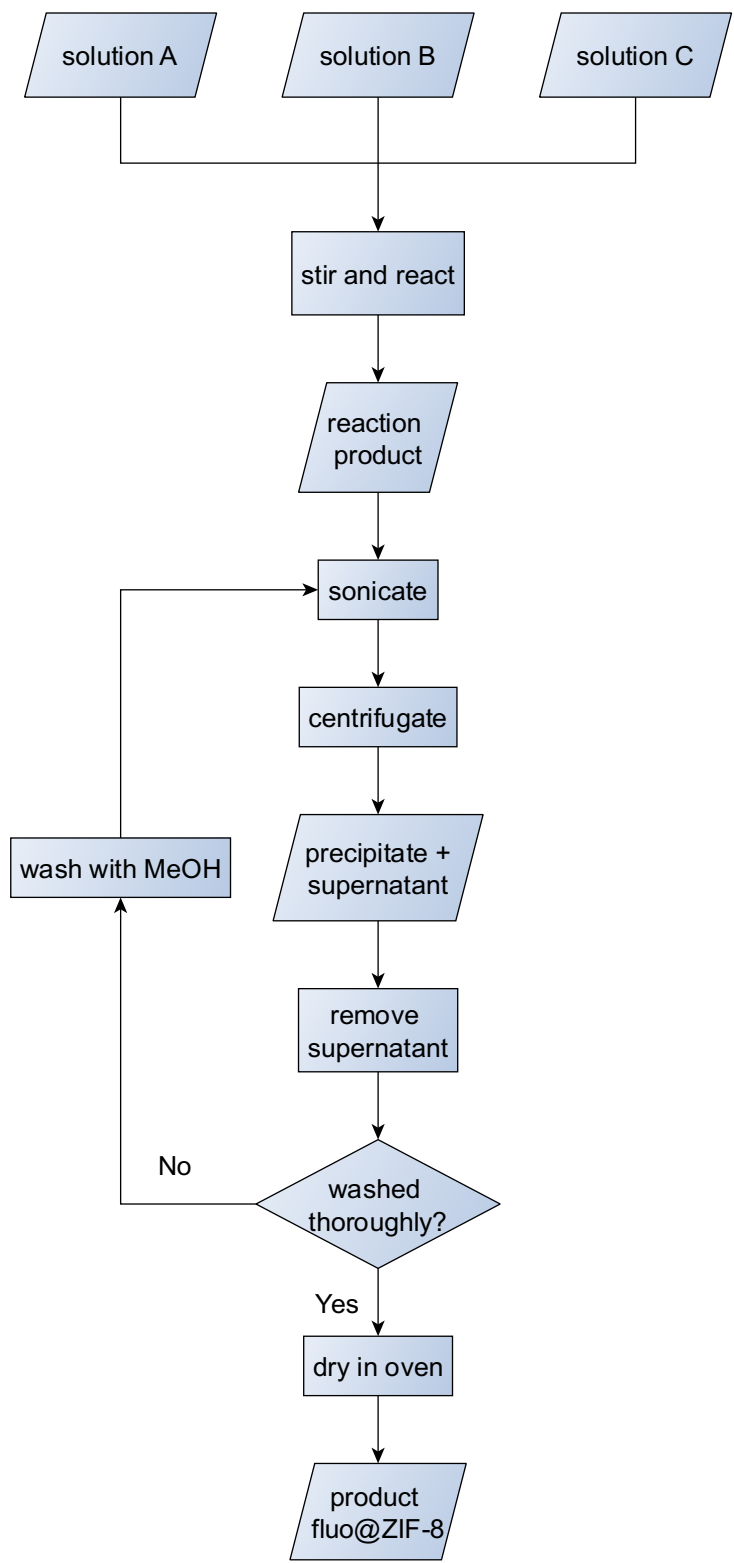

Figure S12. A flowchart of the synthesis of fluo@ZIF-8 samples.

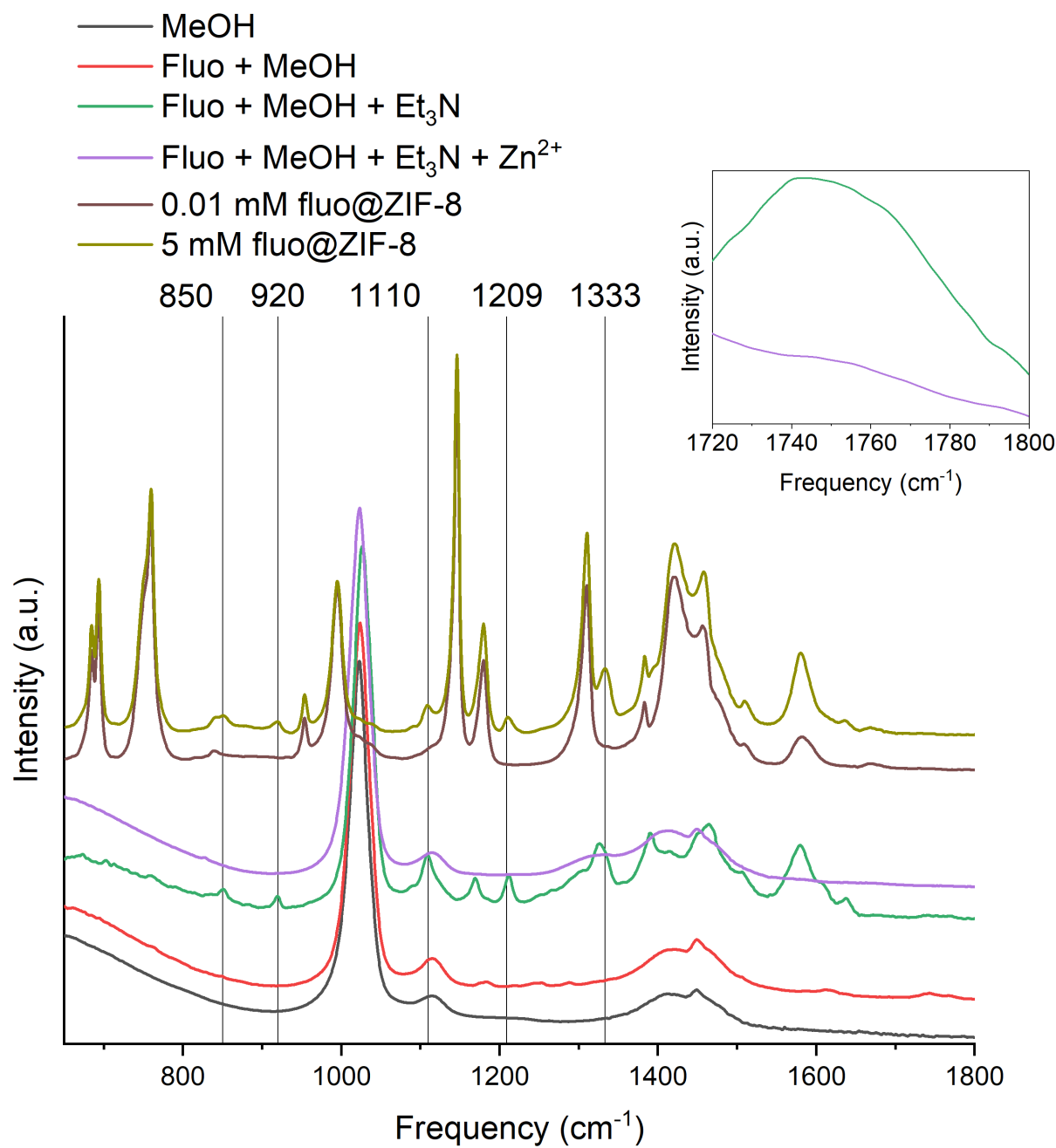

Figure S13. ATR-FTIR spectra for different combinations of solvent (MeOH) and solutes (fluorescein, Et$_3$N, Zn(NO$_3$)$_2$·6H$_2$O), and selected fluo@ZIF-8 samples.

As *ad hoc* comments in response to reviewer questions, we provide the IR spectra of different combinations of solvent (MeOH) and solutes (fluorescein, Et$_3$N, Zn(NO$_3$)$_2$·6H$_2$O), and selected fluo@ZIF-8 samples shown in Figure S13. The concentration of fluorescein in MeOH is about 4.7 mM, determined by trial and error so that reasonably strong FTIR signals for fluorescein can be observed. The molar ratio of fluorescein:Et$_3$N:Zn$^{2+}$ is about 1:15:17.

Note that Et$_3$N is in excess so that fluorescein is deprotonated. The star-marked peaks in Figure 2 of main manuscript (also marked with vertical lines in Figure S13, i.e. 850, 920, 1110, 1209, and 1333 cm$^{-1}$) match the deprotonated fluorescein (Fluo + MeOH + Et$_3$N) peaks very well, justifying the validity of using dianion/anion as the basis of the computation. The inset shows the change of the signal of COO$^-$ group in the region of 1750 cm$^{-1}$, which is a sign that coordination occurred for fluorescein to Zn$^{2+}$. We propose that not all negatively charged fluorescein have Zn$^{2+}$ as counter ions (because the uncoordinated fluorescein seems to fit the experimental data better), but some may be charge balanced by the protonated Et$_3$NH$^+$, the latter case should correspond to the case of the deprotonated fluorescein (Fluo + MeOH + Et$_3$N). With this interpretation, overall we can still observe the new features in 5 mM fluo@ZIF-8 sample that has highest guest loading.

**IR spectra calculated using B3LYP/6-311G* compared with B3LYP/SNSD**

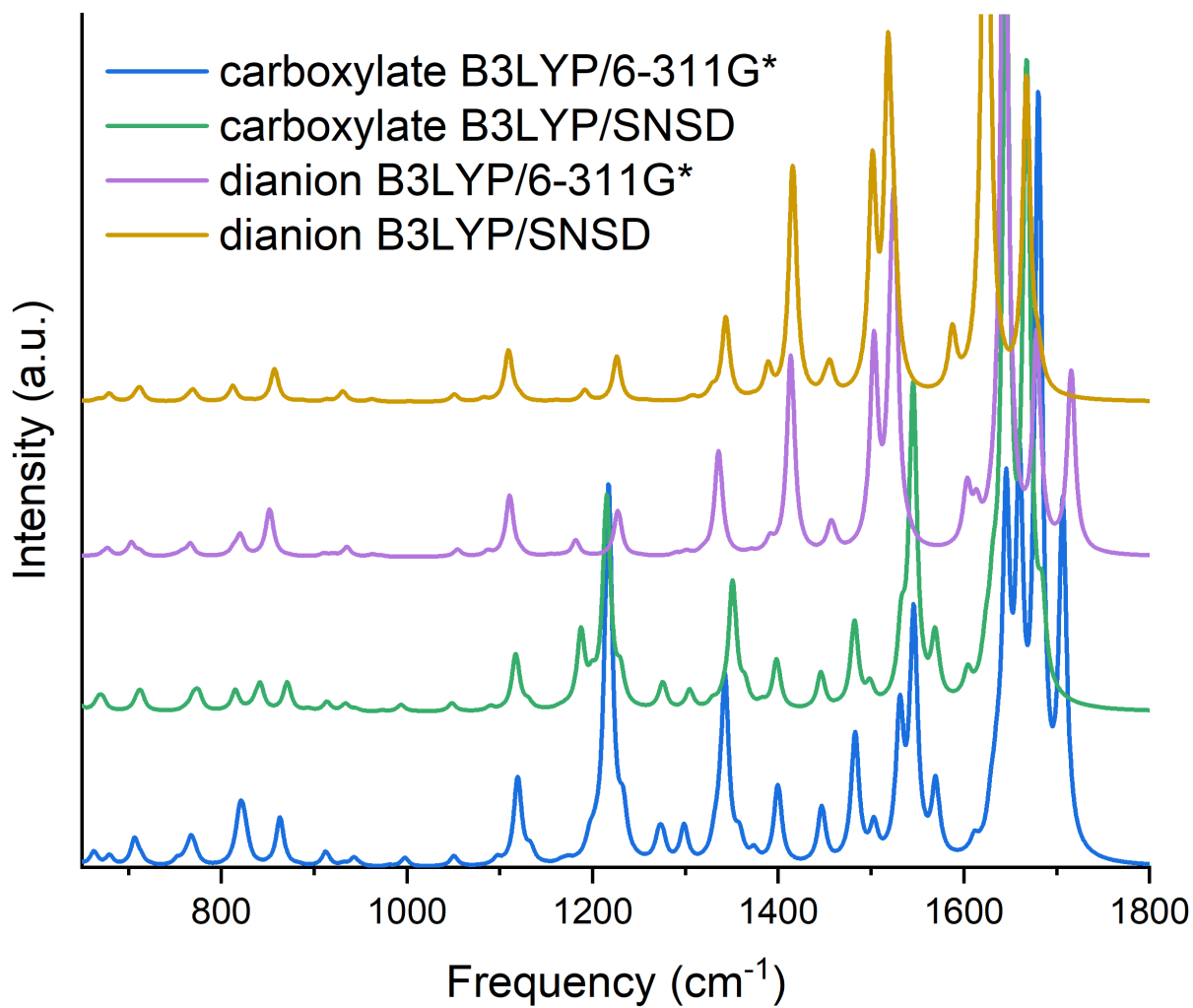

Figure S14. Comparison of IR spectra calculated using B3LYP/6-311G* and B3LYP/SNSD. B3LYP in combination with an improved basis set (SNSD)[6] for IR predictions to calculate the IR spectra of the anion and dianion. A Lorentzian with a FWHM of 10 cm$^{-1}$ is used to broaden the spectra. No scaling factor is used. In regions of interest, no significant deviations are found between the two sets of methods.

**Solution $^1$H NMR spectroscopy of Fluo@ZIF-8**

Samples for NMR were dissolved in a solution composed of 500 μL methanol-d4 and 50 μL DCl / D$_2$O (35 wt%). All NMR spectroscopy was done at 298 K using a Bruker Avance NEO spectrometer operating at 600 MHz, equipped with a BBO cryoprobe. Data was collected using a relaxation delay of 30 s, with 128k points and a sweep width of 19.8 ppm, giving a digital resolution of 0.18 Hz. Data was processed using Bruker Topspin with a line broadening of 1 Hz and 2 rounds of zero-filling. Peaks were integrated using global spectral deconvolution in the MestReNova software package.

The loading amount, defined as the number of fluorescein molecules per cage (of ZIF-8), was calculated from the molar ratio of fluorescein to 2-methylimidazole (mIm). In order to calculate the molar ratio, peaks corresponding to each compound were integrated and normalised according to the number of protons giving rise to the signal. For fluorescein (fluo), the doublet at approximately 8.30 ppm was used, which corresponds to a single proton (that in the ortho position relative to the carboxyl group). For 2-methylimidazole the singlet at approximately 7.31 ppm was used, which corresponds to the two methine protons of the imidazole ring. Global spectral deconvolution (in the MestReNova software package) was used to pick and integrate the peaks. Each ZIF-8 cage contains 12 methylimidazole ligands. The loading amount was therefore calculated as the molar ratio of fluorescein : methylimidazole multiplied by 12.

Table S5. Guest loading determined from solution $^1$H NMR spectroscopy.

| Fluo@ZIF-8 (mM used during synthesis) | Molar ratio | Loading of fluorescein (/100%)* |
|---|---|---|
| 0.01 | 3649.8 | 0.00027 |
| 0.05 | 560.5 | 0.0018 |
| 0.1 | 241.7 | 0.0041 |
| 0.5 | 53.4 | 0.019 |
| 1 | 30.0 | 0.033 |
| 5 | 5.6 | 0.18 |

*The loading amount is defined as the number of fluorescein molecules per ZIF-8 cage.

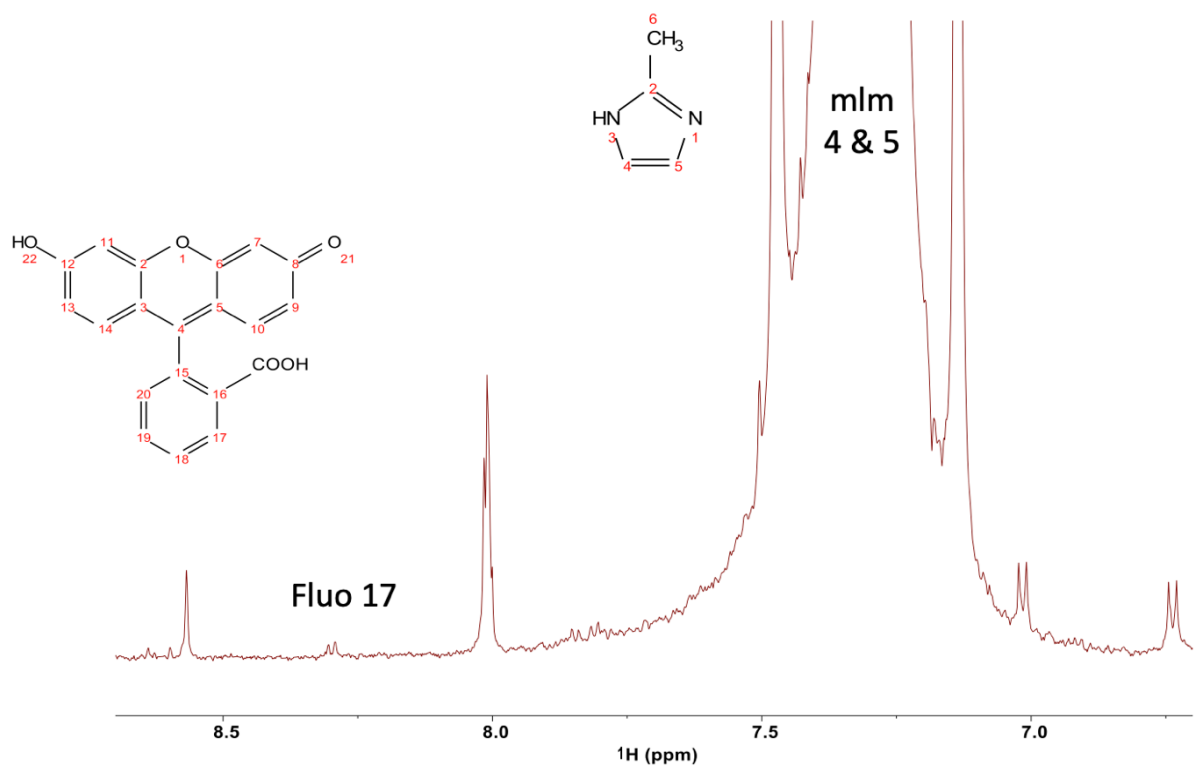

Figure S15. Solution $^1$H NMR of 0.01 mM Fluo@ZIF-8 where the guest/host peaks used for integration are indicated as Fluo and mIm, respectively. The guest loading calculated is 1 fluorescein for every 3703.7 cage.

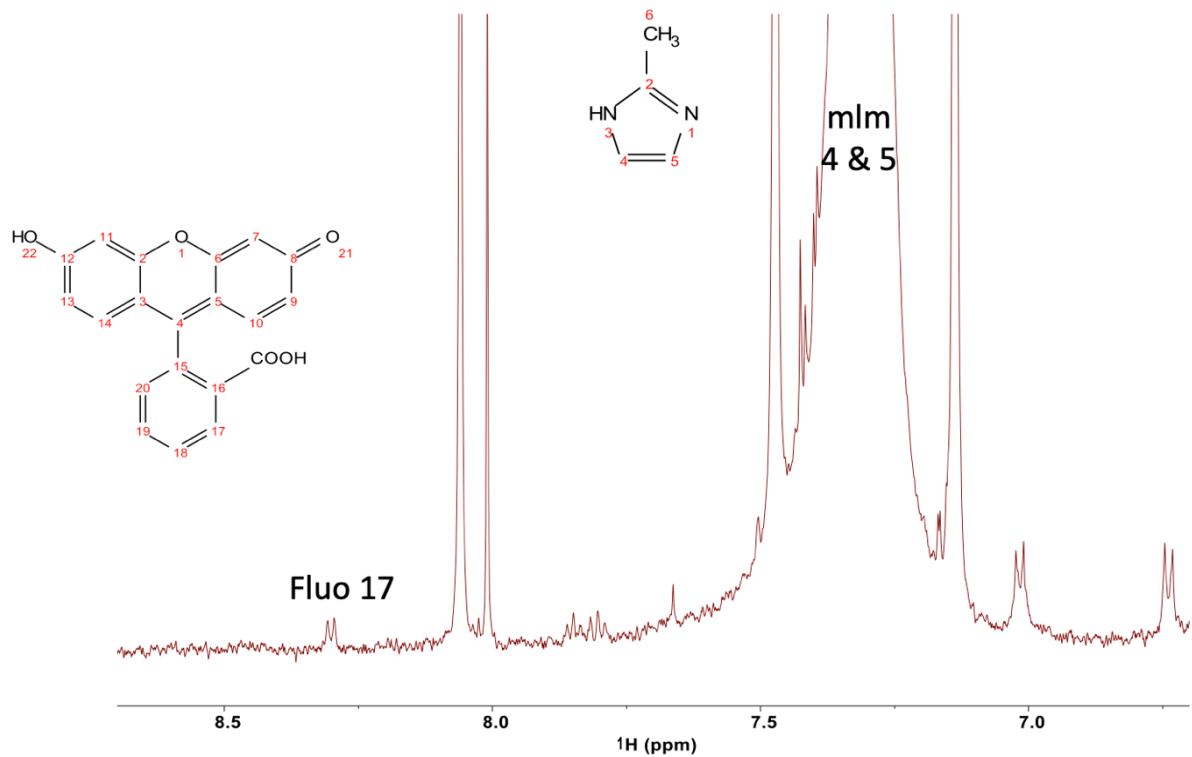

Figure S16. Solution $^1$H NMR of 0.05 mM Fluo@ZIF-8 where the guest/host peaks used for integration are indicated as Fluo and mIm, respectively. The guest loading calculated is 1 fluorescein for every 555.6 cage.

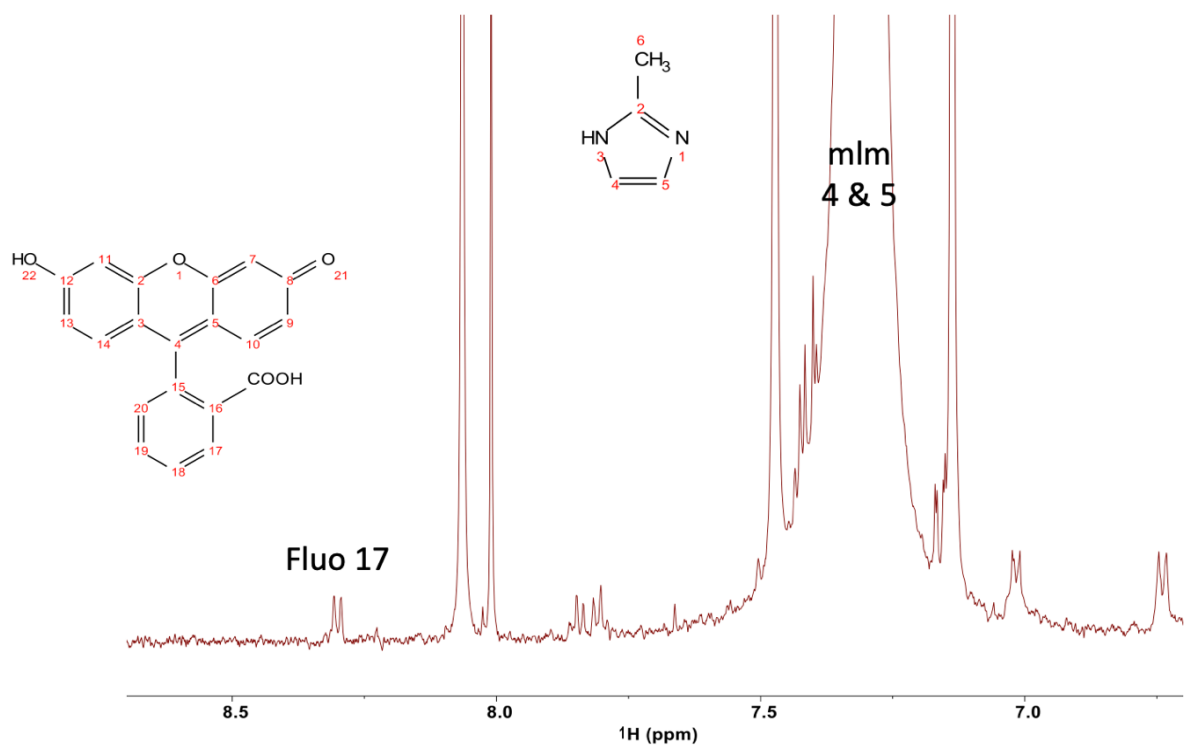

Figure S17. Solution ¹H NMR of 0.1 mM Fluo@ZIF-8 where the guest/host peaks used for integration are indicated as Fluo and mIm, respectively. The guest loading calculated is 1 fluorescein for every 243.9 cage.

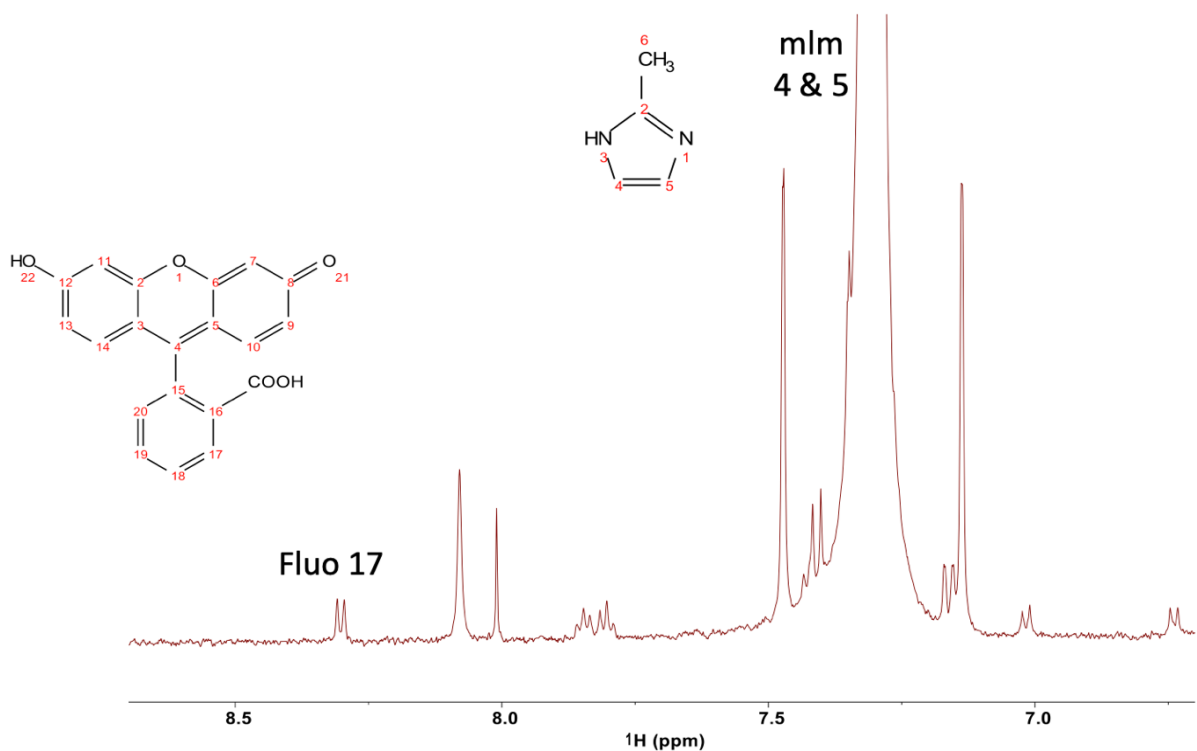

Figure S18. Solution ¹H NMR of 0.5 mM Fluo@ZIF-8 where the guest/host peaks used for integration are indicated as Fluo and mIm, respectively. The guest loading calculated is 1 fluorescein for every 52.6 cage.

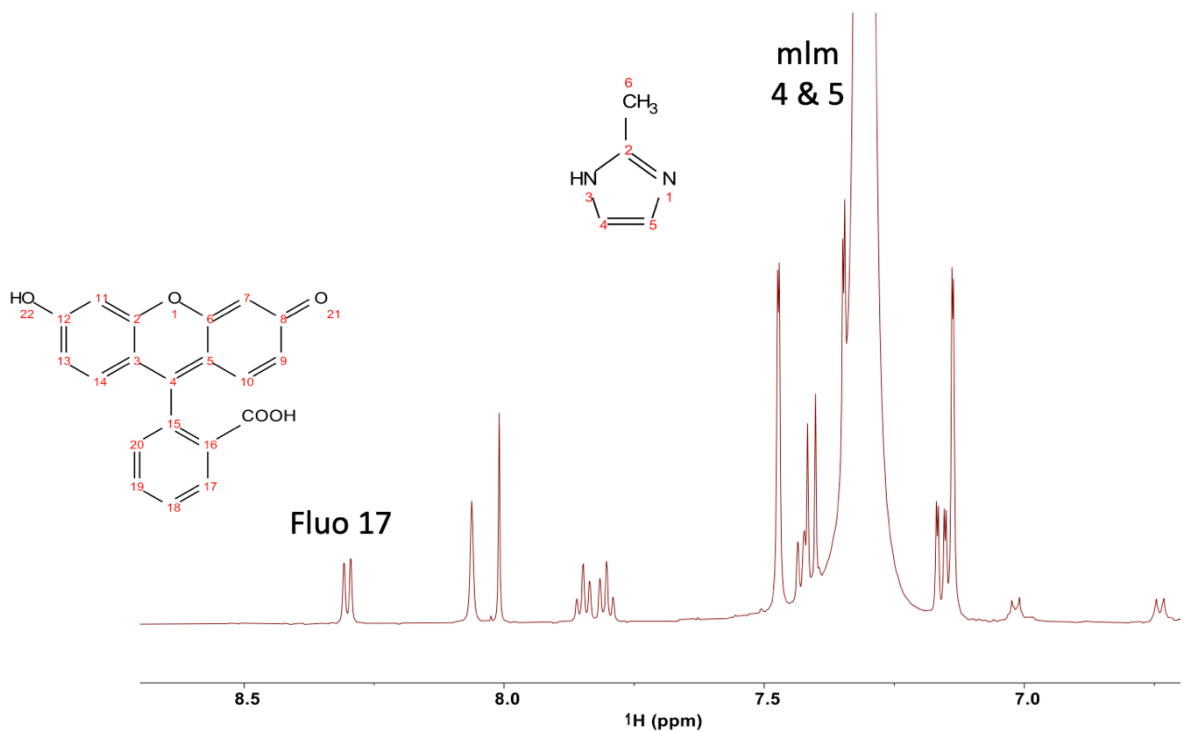

Figure S19. Solution ¹H NMR of 1 mM Fluo@ZIF-8 where the guest/host peaks used for integration are indicated as Fluo and mIm, respectively. The guest loading calculated is 1 fluorescein for every 30.3 cage.

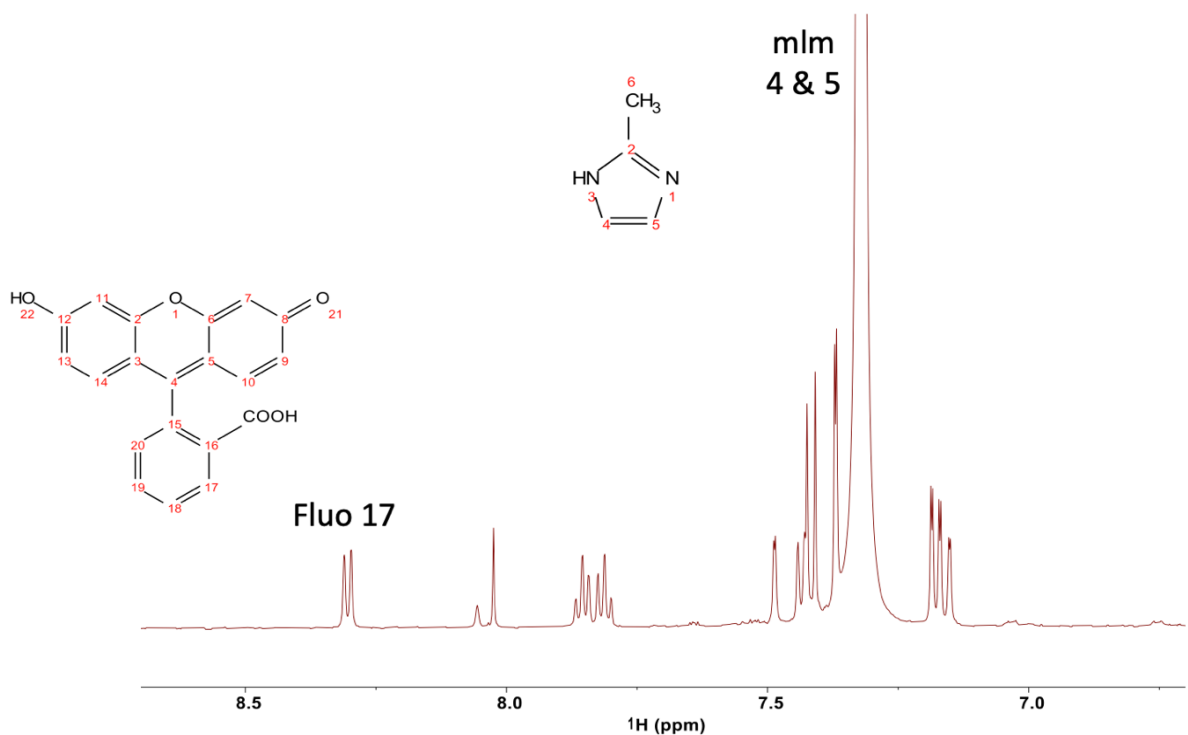

Figure S20. Solution ¹H NMR of 5 mM Fluo@ZIF-8 where the guest/host peaks used for integration are indicated as Fluo and mIm, respectively. The guest loading calculated is 1 fluorescein for every 5.6 cage. Note that it is assumed that all guests are confined in pores even for this high concentration sample.

# References

(1) McIlvaine, T. C. A buffer solution for colorimetric comparison. *J. Bio. Chem.* **1921,** *49*, 183-186.

(2) Sjöback, R.; Nygren, J.; Kubista, M. Absorption and fluorescence properties of fluorescein. *Spectroc. Acta A* **1995,** *51*, L7-L21.

(3) Sun, C.-Y.; Qin, C.; Wang, X.-L.; Yang, G.-S.; Shao, K.-Z.; Lan, Y.-Q.; Su, Z.-M.; Huang, P.; Wang, C.-G.; Wang, E.-B. Zeolitic imidazolate framework-8 as efficient pH-sensitive drug delivery vehicle. *Dalton Trans.* **2012,** *41*, 6906-6909.

(4) Ryder, M. R.; Civalleri, B.; Bennett, T. D.; Henke, S.; Rudić, S.; Cinque, G.; Fernandez-Alonso, F.; Tan, J.-C. Identifying the role of terahertz vibrations in metal-organic frameworks: From gate-opening phenomenon to shear-driven structural destabilization. *Phys. Rev. Lett.* **2014,** *113*, 215502.

(5) Moggach, S. A.; Bennett, T. D.; Cheetham, A. K. The effect of pressure on ZIF-8: Increasing pore size with pressure and the formation of a high-pressure phase at 1.47 GPa. *Angew. Chem. Int. Ed.* **2009,** *48*, 7087-7089.

(6) Barone, V.; Ceselin, G.; Fuse, M.; Tasinato, N. Accuracy meets interpretability for computational spectroscopy by means of hybrid and double-hybrid functionals. *Front Chem* **2020,** *8*, 584203.